\documentstyle[12pt, epsf]{article}
\def\Hpm{ {H^{\pm}}}
\def\h0{{h^0}}
\def\H0{{H^0}}

\newcommand{\lsim}{\raisebox{-0.13cm}{~\shortstack{$<$ \\[-0.07cm] $\sim$}}~}

\begin{document}
\title{\vspace{-15mm}
       \vspace{2.5cm}
       {\normalsize \hfill
       \begin{tabbing}
       \`\begin{tabular}{l}
	 PM/97-25 \\
         UFR-HEP/98-09\\
        August 1998 \\
	\end{tabular}
       \end{tabbing} }
       \vspace{2mm}
Radiative corrections to $e^+ e^- \rightarrow H^+ H^-$: \\ THDM versus MSSM .
}
\vspace{2mm}
 
\author{ A. Arhrib$^{a, b}$\footnote{E-mail:arhrib@fstt.ac.ma}
,  G. Moultaka$^{c}$\footnote{E-mail:moultaka@lpm.univ-montp2.fr}  \\[3mm] 
 {\normalsize \em a: D\'epartement de Math\'ematiques, 
 Facult\'e des Sciences et Techniques,} \\
{\normalsize \em  B.P 416 Tanger Morocco } \\[3mm] 
 {\normalsize \em b: UFR--Physique des Hautes Energies, 
 Facult\'e des Sciences}\\
{\normalsize \em PO Box 1014, Rabat--Morocco}\\[3mm] 
 { \normalsize \em
c: Laboratoire de Physique Math\'ematique et Th\'eorique, CNRS-UMR 5825}\\
{ \normalsize \em 
Universit\'e Montpellier II, F--34095 Montpellier Cedex 5,  France}\\[3mm] }
\setcounter{footnote}{4} 
\date{}
\maketitle 
\begin{abstract}
One loop radiative corrections to $e^+e^-\rightarrow H^+ H^-$
are considered at future linear collider energies, in the general type II
Two Higgs Doublet Model (THDM) and in the Minimal Supersymmetric Standard
Model-like (MSSM) Higgs sector. To make  the  comparison between THDM
and MSSM tractable, we have introduced a quasi--SUSY parameterization which
preserves all the tree-level Higgs mass--sum--rules of the MSSM, 
and involves just 3 free parameters in the Higgs sector
(instead of 7 in the general THDM) and comprises the MSSM as a 
particular case. The model-independent soft photon contribution is
isolated and shown to be substantial. 
Important effects come also from the 
contribution of the model dependent $h^0 H^+ H^-$ and $H^0 H^+ H^-$ vertices
 to the final state.
In the MSSM, the contribution of the Higgs sector is moderate 
(a few percent) while in the THDM and both for small and large tan$\beta$ 
important effects ($\sim +30\%$) can be found. 
\end{abstract}
 
\vfill
\newpage
\section{Introduction}

Supersymmetry is one of the most promising scenarios in terms of which
one can approach new physics at the present and next generation of colliders 
(LEPII, upgraded CDF, LHC,  LC2000,...).
In this context, the awaited discovery of a neutral Higgs particle, 
if successful, 
will need to be complemented by the discovery and study of the 
full-fledged Higgs sector as well as the purely supersymmetric sector, 
in order to test unambiguously the Minimal Supersymmetric Standard Model 
(MSSM) \cite{mssm} or possibly its non-minimal extensions \cite{nmssm}.

Such a task will presumably  be rather intricate
as any realistic supersymmetric model predicts a plethora of 
particles and associated couplings. It should also comprise   
a full test of the various mass-sum-rules and relations among the 
couplings endemic to supersymmetry \cite{peskinetal},
these being, in some cases, sensibly modified by radiative
corrections \cite{19,HempflingHoangQuirosWagner..}.

These large modifications can be crucial in the experimental search strategy.
Perhaps the most illustrative example is the loosening of the MSSM tree-level 
mass bound for the lightest CP-even Higgs
roughly from 
$$ m_{h^0}^2 \leq M_Z^2 \ \ \ \ \ \ \ {\mbox to}\ \ \ \ \ \ \ \
m_{h^0}^2 \lsim M_Z^2 + g^2 \frac{3 m_t^4}{8 \pi^2 M_W^2} \ln 
(\frac{m_{\tilde{t}}^2}{m_t^2} )$$

by loop effects \cite{19,HempflingHoangQuirosWagner..}, 
which raises the mass bound to $\sim 130$ GeV below which
the lightest Higgs should be discovered or else the MSSM ruled out. \\
Sizeable modifications of tree-level predictions are likely to occur
not only in the Higgs mass-sum-rules\cite{sumrule} but also in Higgs 
self-couplings, and more moderately in mass 
relations and couplings in the chargino, neutralino, squark, slepton and
gluino sectors. Some of these have been already looked at \cite{Pierceetal...}
in a model dependent context, however 
ultimately a systematic exploration of all regions of susy parameters
leading to potentially large deviations that can affect tree level relations
as well as production and decay rates of new particles, would become mandatory
in order to achieve full tests of susy. Indeed there is {\sl a priori} no 
guarantee that moderate corrections to a given observable 
will ensure moderate corrections to all other observables related to the same 
given sector.

The subject of the present paper is the one-loop electroweak corrections to 
the charged Higgs pair production in $e^+ e^-$ collisions. 
This will turn out to be precisely an example where important effects 
are present in 
the production cross-section, even though corrections to the $H^\pm$  
mass sum rule are known to be modest 
\cite{20}.
( Loop corrections to $H^\pm$ decays are also important to study 
\cite{Coarasaetal} but will not be considered in this paper).

Generally speaking, a charged scalar particle coupled to fermions proportionally
to their masses is a by-product of an extended Higgs sector with 
scalar fields in  non-singlet representations of the weak symmetry. As such its
experimental discovery would thus be immediate qualitative evidence in favor of
a structure beyond the standard model, in contrast with the neutral Higgs which
would necessitate a more quantitative investigation. Present experimental 
direct search for charged Higgses set lower bound on its mass. At LEP--II 
search,
ALEPH and OPAL Collaborations set a lower bound of 52 GeV and DELPHI 
Collaboration
sets a lower bound of 54.4 GeV \cite{opalcollab}.  At the Tevatron, 
CDF collaboration has
searched for charged Higgs decay of the top quark 
excluding a charged Higgs mass lower than $\sim 147$ GeV in the large 
$\tan\beta$
limit \cite{cdf}. However, 
using the Tevatron quark data in the lepton plus $\tau$ a lower
mass limit of 100 GeV for $\tan\beta\geq 40$ has been reported \cite{roy}.
For a recent discussion of the LEP and Tevatron limits and a reappraisal
see \cite{borzumati} \\

Indirect limits,
valid only in the non supersymmetric case, have been
inferred from $b \to s \gamma$ and read \cite{btosgammanonsusy}: 
$$ M_{H^\pm} > 244 GeV + \frac{63}{\tan \beta^{1.3}} GeV $$
see also \cite{ciuchini}. 

Although supersymmetry can generically invalidate the previous indirect
limit,
some studies suggest that useful bounds can 
still be inferred
in some model-dependent cases (assuming grand unified minimal 
supergravity  and a negative Higgs mass parameter $\mu$ 
\cite{sugra} ).
At hadronic machines, future direct production of charged Higgses
can proceed either via pair production through gluon-gluon fusion \cite{gg}
or the less favoured  $q \bar{q}$ annihilation Drell-Yan process\cite{hadronic}, 
or via single
production in association with  quarks\cite{single} or a $W$ boson\cite{wh}.
On the other hand, the charged Higgs can be efficiently searched for only when 
its top-bottom decay mode is kinematically forbidden
[otherwise one would have to look at bosonic decay modes 
ex. $H^\pm \rightarrow h^0 W^\pm$, $H^\pm \rightarrow A^0 W^\pm$, 
\cite{hadronic}, see also \cite{borzumati}].
 
The future $e^+ e^-$ machines will presumably offer a cleaner environment and in 
that sense 
 a higher mass reach, especially if the 1--2TeV options are available. There 
the charged Higgs can be produced either in pairs through
$\gamma, Z$ exchange\cite{komamiya}, or in single rare production in 
association with a $W$ 
boson \cite{ACMH}. The former channel has at tree-level the particularity of 
depending only
on standard model parameters, it is thus important to understand the trend
of loop effects which contain the model dependence.  

The one-loop electroweak corrections to $e^+e^- \to H^+ H^-$ have been 
previously considered at NLC energies, restricting mainly to top-bottom,
squarks and sleptons contributions \cite{ACM}. In the present paper
we complete the study by including the effects of the remaining 
sectors already summarized in. We also revise some of the numerical results previously given 
in \cite{ACM}.
We mention here incidently that charged Higgs pair production through 
laser back-scattered $\gamma\gamma$ 
collisions  has been studied in the literature at the tree as well 
as the one-loop levels \cite{pp}
reaching similar conclusions as to the importance of the quantum corrections.\\

In view of the potentially large corrections the first issue here will be 
mainly (but not exclusively), to pin down all possible large effects from
other sectors than the ones considered in \cite{ACM}. For this we present the
full set of one-loop corrections in a general type II two-Higgs-doublet
model (THDM-II), that is the complete Higgs sector contributions 
(self-energies, vertices and boxes), the infrared part including initial and 
final soft photon radiation, $\gamma \gamma$, $\gamma Z$ and $W W$
boxes, as well as the (standard) initial state vertex and fermion and gauge
boson self-energies contributions. THDM-II, besides being interesting in its 
own right, is a relevant framework in terms of which to assess the sensitivity
to supersymmetric effects.

The second issue will be to parameterize the comparison between THDM-II
and the more constrained MSSM cases. Indeed it will turn out that apart from 
large model independent contributions in the infrared sector, important
effects are found in diagrams involving triple Higgs couplings, however
only in the non-supersymmetric case. In order to assess the sensitivity
to such effects it will be appropriate to define an effective parameterization 
where all the ( tree-level ) MSSM Higgs mass-sum-rules remain valid while
some of the tree-level triple Higgs couplings deviate from their
MSSM form. This parameterization which we will dub ``quasi-supersymmetric'' 
necessitates just one extra free parameter ( instead of five) in the Higgs
sector, as compared to the MSSM and actually offers a general setting to
quantify the tests of the self-couplings in the Higgs sector. \\

The rest of the paper is organized as follows. In section II we review the main
features of the Higgs potential in THDM-II as well as in the MSSM. We then 
define and discuss the ``quasi-supersymmetric'' parameterization which 
interpolates between the two models and recall the form of the various 
Higgs couplings. In section III the tree-level production 
cross-section is presented and some notations and conventions defined. 
Section IV is devoted to the general form of the radiative corrections and to
a description of the various contributions. In section V we describe the 
renormalization
scheme which is most suited in our case to a comparison between
the supersymmetric and non--supersymmetric cases, and discuss as well the
Infrared sector. Section VI is devoted to the numerical analysis, and
section VII to the conclusion. Finally, presentation of 
complementary analytic expressions and technical details of the calculation
is relegated to the appendices. 
 
\section{ The relevant Lagrangian and the ``quasi-susy'' para\-meterization.}
In this section we focus on the structure of the charged Higgs couplings
to the other Higgs fields and recall as well its couplings to the gauge
bosons and matter fermions, excluding the supersymmetric sector.
We will refrain from presenting extensively the various features of 
the Higgs potential, as this has been repeatedly achieved in the literature
\footnote{we follow throughout this section
the notations of \cite{GunionHaber}  to which we refer the reader for relevant
details} 
The aim here is rather to identify the tree-level couplings which can carry
information about the presence of supersymmetry, distinguishing them from
those which have the same form whether susy is operative or not. We then
encompass these features in a suitable general parameterization.

\subsection{The two-Higgs doublet potential}
Assuming two complex $SU(2)_{weak}$ doublet scalar fields
$\Phi_1$ and $\Phi_2$ defined as
\begin{center}
\[ \Phi_{1}=\left(\begin{array}{c}
\phi^+_{1}=\varphi_{1}+i\varphi_{2}\\ [0.3cm]
\phi^0_{1}=\varphi_{3}+i\varphi_{4}
\end{array}  \right) \hspace{.8in}
\Phi_{2}=\left(\begin{array}{c}
\phi^+_{2}=\varphi_{5}+i\varphi_{6}\\ [0.4cm]
\phi^0_{2}=\varphi_{7}+i\varphi_{8} \end{array}  \right)\]
\end{center} 
the most general (dimension 4) $SU(2)_{weak} \times U(1)_Y$ gauge 
invariant and (CP-invariant) scalar potential is given by
(see, however, the Errata in ref \cite{GunionHaber}):
\begin{eqnarray}
& & V(\Phi_{1},\Phi_{2})=\lambda_{1} (\Phi_{1}^+\Phi_{1}-v_{1}^2)^2
+\lambda_{2} (\Phi_{2}^+\Phi_{2}-v_{2}^2)^2+
                    \lambda_{3}((\Phi_{1}^+\Phi_{1}-v_{1}^2)+(\Phi_{2}^+
\Phi_{2}-v_{2}^2))^2                 \nonumber\\ [0.2cm]
                    & &+\lambda_{4}((\Phi_{1}^+\Phi_{1})(\Phi_{2}^+
\Phi_{2})-(\Phi_{1}^+\Phi_{2})(\Phi_{2}^+\Phi_{1}))+
                    \lambda_{5} (Re(\Phi^+_{1}\Phi_{2})
-v_{1}v_{2})^2+\nonumber\\ [0.2cm]
                    & & \lambda_{6}
Im(\Phi^+_{1}\Phi_{2})^2 +\lambda_{7}
\label{higgspot}
\end{eqnarray}
where $\Phi_1$ and $\Phi_2$ have weak hypercharge Y=1 and the $\lambda_i$'s
 are real-valued. We will also assume 
the arbitrary additive constant $\lambda_{7}$ to be vanishing .
This would of course be the case if exact supersymmetry is imposed, 
and should any way be required for a vanishing cosmological constant (at least
as a tree-level approximation). No other assumptions on the $\lambda_i$'s are
made except that they sit in the regions that allow the required pattern
of spontaneous electroweak symmetry breaking down to $U(1)_{ew}$ \cite{ewsb}.
\\ 
The minimum of the potential is then obtained for 
\[\langle \Phi_{1}\rangle=\left(\begin{array}{c}
v_1\\ [0.2cm]
0
\end{array}  \right) \hspace{.8in}
\langle\Phi_2\rangle=\left(\begin{array}{c}
v_2\\ [0.2cm]
0\end{array}  \right)\]
After the Higgs mechanism has taken place, the W and Z gauge 
bosons acquire masses  
given by  $m_W^2=\frac{1}{2}g^2 v^2$ and $m_Z^2= \frac{1}{2}(g^2 +g'^2) v^2$,
where $g$ and $g'$ are the $SU(2)_{weak}$ and $U(1)_Y$ gauge couplings and
$ v^2= v_1^2 + v_2^2$. The combination $v_1^2 + v_2^2$ 
is thus fixed by the electroweak 
scale through $v_1^2 + v_2^2=(2\sqrt{2} G_F)^{-1}$ 
and we are left with 7 free parameters in eq.(\ref{higgspot}), namely
the $\lambda_i$'s and $v_2/v_1$.

On the other hand, three of the eight 
degrees of freedom of the two Higgs doublets correspond to the 3 
goldstone bosons, the remaining five become the physical Higgs bosons
$H^0, h^0$ (CP-even), $A^0$ (CP-odd) and $H^\pm$. 
Their masses are obtained as usual from the shift
$ \Phi_i \to \Phi_i + \langle \Phi_i \rangle$ and read     
\begin{eqnarray}
&& m_{A^0}^2=\lambda_6 v^2\  ;\ \ \ \  m_{H^{\pm}}^2=
\lambda_4 v^2\ \ \ \mbox{and} \ \
\ \ m_{H^0,h^0}^2=\frac{1}{2} [ A+C \pm \sqrt{(A-C)^2+4B^2} ] \label{higgsmass}
\end{eqnarray}
where
\begin{eqnarray}
&& A=4 v_1^2 (\lambda_1+\lambda_3)+v_2^2\lambda_5\ , \ \  B= v_1 v_2
(4\lambda_3+\lambda_5)\ \ and \ \ \
C=4 v_2^2 (\lambda_2+\lambda_3)+v_1^2\lambda_5 \label{ABC}
\end{eqnarray}
The angle $\beta$ given by $\tan \beta= v_2/v_1$ defines the mixing leading to
the physical $H^\pm$ and $A^0$ states, while the mixing angle $\alpha$ 
associated to $H^0, h^0$ physical states is given by 
[See ref \cite{GunionHaber} for further details.]
\begin{eqnarray}
&& \sin 2 \alpha=\frac{2 B}{\sqrt{(A-C)^2+4 B^2} }\ , \ \ \
\cos 2 \alpha=\frac{A-C}{\sqrt{(A-C)^2+4 B^2} }
\label{alph}
\end{eqnarray}
It will be more suitable for the forthcoming discussion to trade the five
parameters $\lambda_{1, 2,4,5,6}$ for the 4 Higgs masses 
and the mixing angle $\alpha$. From now on we will take the physical Higgs masses,
$m_{H^0}, m_{h^0}, m_{A^0}, m_{H^\pm} $, the mixing angles $\alpha, \beta$ and
the coupling $\lambda_3$ as the 7 free parameters. 

It is then straightforward algebra to invert equations (\ref{higgsmass})
through (\ref{alph}), and get the $\lambda_i$'s in terms of
this new set of parameters.   
\begin{eqnarray}
& & \lambda_4=\frac{g^2}{2 m^2_W} m_{H^\pm}^2 \ \ \ \ \ \ \ \ \ \ \ \ \
\lambda_6= \frac{g^2}{2 m^2_W} m_{A}^2  \label{lambda46} \\ & &
\lambda_5=\frac{g^2}{2 m^2_W} \frac{\sin 2\alpha}{ \sin 2 \beta} (m_H^2-m_h^2)\ -
\ 4 \lambda_3 \label{lambda5} \\ 
&&\lambda_1= \frac{g^2}{16 \cos^2\beta m^2_W} [m^2_H+m^2_h +
(m^2_H - m^2_h)\frac{\cos(2\alpha +\beta)}{\cos\beta}]
+ \lambda_3(-1 + \tan^2\beta) \label{lambda1}  \\
&& \lambda_2= \frac{g^2}{16 \sin^2\beta m^2_W} [m^2_H+m^2_h +
(m^2_h - m^2_H)\frac{\sin(2\alpha +\beta)}{\sin\beta}]
+ \lambda_3(-1 + \cot^2\beta) \label{lambda2}
\end{eqnarray}
Using the above expressions, and after having expressed 
eq.(\ref{higgspot}) in terms of the Goldstone and physical Higgs fields,  
one can cast all couplings of the scalar potential
in terms of the new set of free parameters.\\  
Here we restrict
ourselves to all 3-point vertices (see eqs. (\ref{54}--\ref{58}) of appendix A)
 involving a charged Higgs field, since these
are the only ones that enter one-loop calculations in 
$e^+e^- \to H^+ H^-$ (4-point vertices involving $H^\pm$ have vanishing 
contributions). Injecting eq.(\ref{lambda46}--\ref{lambda2}) in eqs. 
(\ref{54}--\ref{58})
we get the various couplings in terms of the new set of free parameters,
\begin{eqnarray}
g_{H^0H^+H^-}& = &-i\frac{g}{m_W} [\cos(\beta-\alpha)(m^2_{\Hpm}
- \frac{m^2_H}{2}) + \frac{\sin(\alpha+\beta)}{sin 2\beta}\{
4\lambda_3 v^2 +\frac{1}{2}(m^2_H+m^2_h)\nonumber \\
&&-\frac{1}
{2 \sin 2\beta}(\sin 2\alpha + 2 \sin (\alpha-\beta)
\cos(\alpha +\beta))(m^2_H-m^2_h) \}] \\
g_{h^0H^+H^-}&=& -i\frac{g}{m_W} [\sin(\beta-\alpha)(m^2_{\Hpm}
- \frac{m^2_h}{2}) + \frac{\cos(\alpha+\beta)}{\sin 2\beta}\{
4\lambda_3 v^2 +\frac{1}{2}(m^2_H+m^2_h) \nonumber \\
&&-\frac{1}
{2 \sin 2\beta}(\sin 2\alpha + 2 \sin (\alpha+\beta)
\cos(\alpha -\beta))(m^2_H-m^2_h) \}]  \\
g_{H^0 H^{\pm} G^{\mp}}&=&
=\frac{-i g \sin(\beta-\alpha) (m_{H^\pm}^2-m_H^2)}{2 m_W}\\
g_{h^0 H^{\pm} G^{\mp}}&=&
\frac{i g \cos(\beta-\alpha) (m_{H^\pm}^2-m_h^2)}{2 m_W}\\
g_{A^0 H^{\pm} G^{\mp}}&=& \mp \frac{v}{\sqrt{2}}\ (\lambda_6-\lambda_4)=
\mp \frac{m_{H^\pm}^2-m_A^2}{v \sqrt{2}}
\end{eqnarray}
It is noteworthy that $\lambda_3$ enters only $g_{H^0H^+H^-}$
and $g_{h^0H^+H^-}$ while $g_{H^0 H^{\pm}G^{\mp}}, g_{h^0 H^{\pm}G^{\mp}}$
and $g_{A^0 H^{\pm}G^{\mp}}$ have automatically their MSSM form, however,
without assuming the MSSM mass-sum-rules. 
As we will see in the next subsection, a direct consequence when MSSM
mass-sum-rules are assumed, will be that $\lambda_3$ measures the ``amount''
of ``hard'' susy breaking, and that among the five couplings, only 
$g_{H^0H^+H^-}$ and  $g_{h^0H^+H^-}$ are sensitive to such a breaking. 
We should also stress that
no constraint has been imposed thus far, in the derivation of the preceding
expressions. In particular no sign assignments for the trigonometric functions
of $\alpha$ or $\beta$ where made ( apart of course from the definition
$\tan \beta \equiv v_2/v_1 > 0$)
\subsection{The ``quasi-susy'' parameterization}
We proceed hereafter to the definition of a partly constrained
parameterization which interpolates between the free THDM case, and the
constrained MSSM, in the following sense:
\begin{itemize}
\item[a)] It preserves {\it all} tree-level mass-sum-rules of the MSSM,
including the relation between $\tan 2 \alpha$ and $\tan 2 \beta$
\footnote{quantum corrections should be consistently omitted here,
 as they contribute non-leading effects to the one-loop corrected
$e^+e^- \to H^+ H^-$, see also discussion in section 5};
\item[b)] It needs just 3 free parameters (instead of 7 ) and comprises the
susy case as a special case with 2 free parameters;
\end{itemize}
Actually such a parameterization would be phenomenologically very natural in 
case where 
more than one Higgs particle is discovered, with masses consistent with
the MSSM sum-rules. Indeed, taking these sum-rules as granted, 
one can devise a strategy for further precision 
tests of the supersymmetric self-Higgs couplings, 
where a reduced number of free parameters is involved 
and particularly sensitive sectors are identified .  

Let us first recall the situation when softly broken supersymmetry is 
imposed in eq.(1). As shown in \cite{GunionHaber} one has the following
relations among the $\lambda_i$'s:       
\begin{eqnarray}
&&\lambda_1= \lambda_2  \qquad , \qquad
\lambda_3=  \frac{1}{8}(g^2+g'^2)-\lambda_1 \qquad , \qquad
 \lambda_4= 2 \lambda_1 -\frac{1}{2}g'^2  \nonumber \\ & &
\lambda_5=  -\frac{1}{2}(g^2+g'^2)+ 2 \lambda_1\qquad , \qquad
 \lambda_6=  -\frac{1}{2}(g^2+g'^2)+ 2 \lambda_1
\end{eqnarray}
Due to these five constraints one is left with two free parameters, ex. $\tan
\beta $ and $m_A$. It should be stressed that the soft breaking, in the Higgs 
sector, is actually
encoded in this two-parameter freedom. (Indeed, if exact susy were
 imposed, $V_{soft}=0$, 
then all $\lambda_i$'s would be fixed in terms of the gauge couplings 
$g, g'$, together
with $\tan \beta =1$ and $m_A=0$ \cite{GunionHaber} ). 
Subsequently, the presence (at the tree-level) of more than these two free 
parameters in the Higgs 
sector would be associated to the hard breaking of supersymmetry.
The most general parameterization of such a breaking leading back to the general
THDM discussed in the previous section, can be cast in the form: 
\begin{eqnarray}
&&\lambda_1= \lambda_2 +\delta_{12} \ \ \ \ \ \ \ \ \ \ ,\ \ \ \ \ \ \ \ \
\lambda_3=  \frac{1}{8}(g^2+g'^2)-\lambda_1 +\delta_{31} \nonumber\\
&& \lambda_4= 2 \lambda_1 -\frac{1}{2}g'^2 + \delta_{41}  \ \ \ \ \ , \
\ \ \ \lambda_5=  -\frac{1}{2}(g^2+g'^2)+ 2 \lambda_1  +
\delta_{51}  \nonumber \\
&&\lambda_6=  -\frac{1}{2}(g^2+g'^2)+ 2 \lambda_1 + \delta_{61} 
\label{hardbreaking}
\end{eqnarray}
where the $\delta$'s measure the amount of hard breaking of supersymmetry.
 
The various Higgs masses and $\tan 2 \alpha$ can then be related to the
$\delta$'s upon use of the above equations in conjunction with
 eq.(\ref{higgsmass}--\ref{alph}) as, 
\begin{eqnarray}
& & tan2\alpha = tan2\beta
\frac{m^2_A+m^2_Z -v^2 (\delta_{51}+\delta_{61}+ 4 \delta_{31})}
{m^2_A -m^2_Z -v^2(-\delta_{51}+\delta_{61}+ 4\delta_{31}
+4 \frac{tan^2 \beta}{1-tan^2 \beta} \delta_{12})}\nonumber\\
& &m^2_{H^\pm}= m^2_A+m^2_W +v^2(\delta_{41} - \delta_{61})\\
& &m^2_H +m^2_h =
m^2_A + m^2_Z +v^2 ( 4\delta_{31}+\delta_{51}
-\delta_{61} -4 \sin^2\beta \delta_{12} ) \nonumber\\
& & m^2_H- m^2_h = \sqrt{(\cos 2 \beta (m^2_Z - m^2_A +
 v^2(-\delta_{51}+\delta_{61}+ 4\delta_{31}+4 \frac{tan^2 \beta}{1-tan^2 
\beta} \delta_{12} ) )
)^2 (1 + tan^2 2 \alpha)}\nonumber
\label{massrelations}
\end{eqnarray}
Now the crucial point is to note that the MSSM tree-level mass-sum-rules,
\begin{equation}
m^2_{H^0, h^0}= \frac{1}{2} (m^2_Z + m^2_A \pm \sqrt{(m^2_Z+ m^2_A)^2 -
             4 m^2_Z m^2_A \cos^2 2\beta}) \label{neutralmass}
\end{equation}
\begin{equation}
m^2_{H^\pm} = m^2_A + m^2_W
\label{chargedmass}
\end{equation}
as well as
\begin{equation}
tan 2\alpha = tan 2\beta \frac{m^2_A+m^2_Z}{m^2_A -m^2_Z}
\label{alphabeta}
\end{equation}
can be recovered {\sl not only in the susy case $\delta's=0$, 
but actually in a full one-parameter subspace}. 
Indeed the validity of 
eq.( \ref{neutralmass}-\ref{alphabeta} ) requires the following equations
\begin{eqnarray}
& &\delta_{51}+\delta_{61}+ 4 \delta_{31} = 0 \qquad , \qquad
 \delta_{61} - \delta_{51} + 4\delta_{31}
+4 \frac{tan^2 \beta}{1-tan^2 \beta} \delta_{12} = 0 \nonumber \\
& & \delta_{41} - \delta_{61} = 0 \qquad \qquad \quad , \qquad
 4\delta_{31}+\delta_{51}
-\delta_{61} -4 \sin^2\beta \delta_{12} = 0 
\label{deltaeq}
\end{eqnarray}
which are satisfied if
\begin{eqnarray}
&&\delta_{12}= \frac{( tan^4 \beta - 1)}{tan^4 \beta} \delta_{31} \qquad , 
\qquad \delta_{51}= \frac{ -2 ( 1 + tan^2 \beta)}{tan^2 \beta} \delta_{31} \nonumber \\
&& \delta_{61}=\delta_{41}= \frac{ 2 ( 1 - tan^2 \beta)}{tan^2 \beta} 
\delta_{31} 
\label{deltasol}
\end{eqnarray}
With the above four constraints, we have just one extra free parameter as 
compared to the MSSM case,
ex. $\delta_{31}$ or equivalently $\lambda_3$ eq.(\ref{hardbreaking}) on top of 
$m_A$ and $\tan \beta$. 
Eqs.(\ref{deltasol}) together with the set of free parameters
($\lambda_3, m_A, \tan \beta$), define our quasi-susy (QSUSY) parameterization. 
Obviously, when $\lambda_3$ hits its supersymmetric value, i.e. when $\delta_{31}=0$,
we get back the MSSM case, otherwise $\Delta \lambda_3= \lambda_3 -\lambda^{susy}_3$
can be seen as measuring the ``hardness'' of the susy breaking, while $m_A$ and 
$\tan \beta$
play the same role as in the MSSM.

There is a subtlety, however, as concerns the sign of $\sin 2 \alpha$. 
Eqs.(\ref{deltasol})
are sufficient to lead to a looser constraint as compared to the well-known 
supersymmetric constraint
\cite{habergunion2} $\sin 2 \alpha <0$. Actually one can show 
that\footnote{ Detailed derivation
of these and subsequent results related to QSUSY parameterization 
in the full Higgs
sector including the leading one-loop corrections, will be given elsewhere}
\begin{itemize}
\item[a)] $\sin 2 \alpha <0$ can be chosen in all the parameter space;
\item[b)] $\sin 2 \alpha >0$ is possible provided $\cos 2 \alpha <0$, $\tan \beta <1$
and $ m_A < m_Z$;
\end{itemize}
Case b) means that the only consistent
choice becomes  $\sin 2 \alpha <0$ as soon as $\tan \beta >1$ or $ m_A > m_Z$. 
Since in our study we assume a very heavy
charged Higgs ($ m_{H_\pm} > 2 m_Z$), thus a very heavy CP-odd neutral Higgs,
 eq.(\ref{chargedmass} ), 
we will stick to the choice $\sin  2 \alpha<0$ 
throughout the paper. 
One can then determine uniquely the behavior of the various couplings
in terms of $\lambda_3, m_A, \tan \beta$:
\begin{eqnarray}
g_{H^0H^+H^-}& =  g_{H^0H^+H^-}^{MSSM} - i g m_W 
(\frac{1}{2 c_w^2}+\frac{m_A^2}{m_W^2}+
\frac{s_w^2}{\pi \alpha} \lambda_3)\tan \beta \nonumber \\
g_{h^0H^+H^-}& =  g_{h^0H^+H^-}^{MSSM} - i g m_W \frac{2 m_A^2}{m_A^2-m_Z^2} 
(\frac{1}{2 c_w^2}+\frac{m_A^2}{m_W^2}+
\frac{s_w^2}{\pi \alpha} \lambda_3) 
\label{largetbeta}
\end{eqnarray}
where
\begin{eqnarray}
g_{H^0H^+H^-}^{MSSM}& =& -i g (m_W \cos(\beta - \alpha) - \frac{m_Z}{2 c_w} 
\cos 2 \beta \cos(\beta + \alpha)) \nonumber \\
g_{h^0H^+H^-}^{MSSM}& =& -i g (m_W \sin(\beta - \alpha) + \frac{m_Z}{2 c_w} 
\cos 2 \beta \sin(\beta + \alpha)) \nonumber \\
g_{H^0 H^{\pm}G^{\mp}}&=&
\frac{-i g \sin(\beta-\alpha) (m_{H^\pm}^2-m_H^2)}{2 m_W}\nonumber\\
g_{h^0 H^{\pm}G^{\mp}}&=&
\frac{i g \cos(\beta-\alpha) (m_{H^\pm}^2-m_h^2)}{2 m_W}\nonumber\\
g_{A^0 H^{\pm}G^{\mp}}&=& 
\mp \frac{m_{H^\pm}^2-m_A^2}{2 m_W} 
\label{gH}
\end{eqnarray} 
\subsection{Charged Higgs-bosons interactions with gauge-bosons}
The Higgs-bosons interactions with gauge-bosons are model independent.
These interactions arise from the covariant derivatives in the Lagrangian:
\begin{eqnarray}
\sum_i (D_{\mu}\Phi_i)^+(D_{\mu}\Phi_i)=\sum_i [(\partial_\mu +ig \vec{T_a} 
\vec{W_{\mu}^a}  +ig'\frac{Y_{\Phi_i}}{2}B_\mu) \Phi_i]^+(\partial_\mu +ig 
\vec{T_a} \vec{W_{\mu}^a} +ig'\frac{Y_{\Phi_i}}{2}B_\mu )\Phi_i \label{24} 
\end{eqnarray}
where: $\vec{T_a}$ are the isospin gauge generators, $Y_{\Phi_i}$ 
the hypercharge of the
 Higgs fields,  ${W^a}_\mu$ the $SU(2)_L$ gauge fields,
$B_\mu$ the $U(1)_Y$ gauge field, and $g$ (resp. $g'$) the associated coupling
constants.\\ From eq. (\ref{24}), one can easily extract the coupling of 
charged Higgs pair to 
the photon $A_\mu$, and Z boson $Z_\mu$. The corresponding Feynman rules read,
\begin{eqnarray}
& & A_\mu  H^+H^-=-ie (k_1-k_2)_\mu   \qquad  , \qquad
 Z_{\mu}\ H^+H^-=-i e \frac{c_W^2-s_W^2}{2 s_W c_W}  (k1-k_2)_\mu
\end{eqnarray} 
Where $k_{1,2}$ are the incoming momentum of the charged Higgs boson.
Note that these vertices depend only on standard parameters 
(the electric  charge and Weinberg angle ). 
At tree--level, the $H^\pm W^\mp A_\mu$ vertex 
does not exist as a consequence of the
conservation of the
electromagnetic current. On the contrary the
vanishing of the $H^\pm W^\mp Z_\mu$ vertex is accidental, (see \cite{18}
for further discussions). 
A consequence of
the absence of these
two vertices at tree-level is that the mixing $H^\pm$--$W^\mp $ is not present in our
study. From eq. (24) we can get also all the Feynman rules for the three and four point
vertices involving Higgs and gauge bosons.
Obviously all these Feynman rules have of the same structure in any THDM,
irrespective of the implementation of supersymmetry,
and depend only on the mixing angles $\alpha$ and $\beta$ (Appendix C). 
The only impact of SUSY there, as compared to a general THDM model, would be
to require  $\sin 2\alpha \le 0$.
  
\subsection{Charged Higgs-bosons interactions with fermions}
In the two Higgs doublets extension of the standard model, there exist
two different ways to couple Higgs fields to matter:
either {\sf type I} where the quarks and leptons couple
exclusively to one of the two Higgs doublets,
exactly as in the minimal standard model and will not be considered further
in this paper, or   
the {\sf type II} where, to avoid the problem of Flavor Changing
Neutral Current (FCNC) \cite{glashow-weinberg},
$\Phi_1$ couples only to down quarks (and charged leptons)
and $\Phi_2$
couples only to up quarks (and neutral leptons). 
This latter model
is the pattern found in the MSSM. 
In this case, the charged Higgs interaction to fermions is given by: 
\begin{equation}
H^- u\bar d=\frac{g V_{ud} }{ \sqrt{2} m_W} \Bigm ( Y_u
\frac{(1-\gamma_5)}{2} + Y_d \frac{(1+\gamma_5)}{2}
\Bigm )\label{27}
\end{equation}
where $Y_u=\frac{m_u}{tan\beta} $ and $Y_d=tan\beta\ m_d $, 
$V_{ud}$ is the Kobayashi--Maskawa matrix element which we will take close to 
one.
 
\section{Notations, conventions and cross section in the lowest order.}
In this paper we will use the following notations and conventions.
 The momentum of the outgoing electron and positron and incoming
Higgs bosons $H^+,\ H^-$ are denoted by $p_1$, $p_2$, $k_1$ and $k_2$,
respectively. Neglecting the electron mass $m_e$ (and also the
electron--Higgs couplings which is proportional to $m_e$),
the momenta in the center--of--mass system of the $e^+e^-$ are given by:
\begin{eqnarray}
& & p_{1,2}=\frac{\sqrt{s}}{2} (1,0,0,\pm 1) \nonumber \\
& & k_{1,2}=\frac{\sqrt{s}}{2} (1,\pm \kappa \sin\theta,0,\pm \kappa
cos\theta) \nonumber
\end{eqnarray}
where $\sqrt{s}/2$ denotes the beam energy, $\theta$ the scattering angle
between the $e^+$ and $H^+$ flight directions in the laboratory frame, 
$\kappa ^2=1-\frac{\textstyle 4 m_{H^\pm}^2}{\textstyle s}$, and $m_{H^\pm}$
the mass of the charged Higgs.\\
The Mandelstam variables are defined as follow:
\begin{eqnarray}
& & s =  (p_1+p_2)^2 = (k_1+k_2)^2  \nonumber\\
& & t = (p_1-k_1)^2 = (p_2-k_2)^2 = m_{H^+}^2- \frac{s}{2}
+\frac{s}{2} \kappa \cos\theta  \nonumber\\
& & u = (p_1-k_2)^2 = (p_2-k_1)^2 = m_{H^+}^2- \frac{s}{2}-
\frac{s}{2} \kappa \cos\theta \nonumber \\
& & s+t+u = 2 m_{H^+}^2  \nonumber
\end{eqnarray}
The only Feynman 
diagrams contributing at the tree-level are the $\gamma$ and $Z$ s-channel
exchange, [the electron--Higgs coupling being negligibly small, there is no t-channel
contribution at this order.]
\def\fe#1#2{
\begin{picture}(100,200)(0,0)
\put(34,0){\circle*{5}}
\put(0,35){\vector(1,-1){35}}
\put(-10,26){\makebox(0,0){#1}}
\put(-10,-26){\makebox(0,0){#2}}
\put(0,-35){\vector(1,1){35}}
\end{picture} }
%
\def\bp#1#2{
\begin{picture}(100,200)(0,0)
\multiput(0,0)(16,0){3}{\oval(8,8)[t]}
\multiput(8,0)(16,0){3}{\oval(8,8)[b]}
\put(#2,15){\makebox(0,0){#1}}
\end{picture}   }
\def\fsp#1#2{
\begin{picture}(100,200)(0,0)
\put(0,0){\circle*{5}}
\multiput(0,0)(16,16){3}{\line(1,1){10}}
\multiput(0,0)(16,-16){3}{\line(1,-1){10}}
\put(27,36){\makebox(0,0){#1}}
\put(52,-36){\makebox(0,0){#2}}
\end{picture}   }
\begin{center}
\begin{picture}(100,100)(0,0)
\put(0,50) {\fe{$e^+$}{$e^-$}}
\put(40,50) {\bp{$\gamma,Z$}{24}}
\put(85,50) {\fsp{$H^+$}{$H^-$} }
\end{picture}
\end{center}
\vspace{.4cm}
The contributions to the lowest-order amplitude have the following form:
\begin{eqnarray}
{\cal M}_0^{\gamma} &=& -2\frac{e^2}{s}
\bar {v}(p_2)\not k_1 u(p_1)\label{2.28}\\
{\cal M}_0^{Z}  &=& 2 \frac{e^2g_H}{s-m_z^2} \bigm ( g_V \bar {v}(p_2)
\not k_1 u(p_1) -
   g_A \bar {v}(p_2) \not k_1 \gamma ^5 u(p_1) \bigm )\label{2.29}
\end{eqnarray}
where $g_V=(1-4 s_W^2)/(4 c_W s_W)$, $g_A=1/(4 c_W s_W)$,
$g_H=-(c_W^2-s_W^2)/(2 c_W s_W)$, $c_W\equiv\cos\,\theta_W$,
$s_W\equiv \sin\, \theta_W$.\\
As one can see from eq. (\ref{2.28}, \ref{2.29}), 
those amplitudes can be expressed in terms of 
two invariants, $I_V$ and $I_A$ 
defined as:
\begin{equation}
I_V=\bar {v}(p_2)\not k_1 u(p_1), \ \ \ \ \ \ \ \  
I_A=\bar {v}(p_2) \not k_1 \gamma^5 u(p_1)\label{29}
\end{equation}
The Born amplitude is given by:
\begin{equation}
{\cal M}_{0}={\cal M}_0^{\gamma}+ {\cal M}_0^{Z}\label{30}
\end{equation} From eqs. (\ref{2.28}, \ref{2.29}), the corresponding 
differential cross section 
first studied in \cite{komamiya} is found to be, 
\begin{eqnarray}
\left(\frac{d \sigma}{d \Omega}\right)_0=\frac{\alpha^2\kappa^3}{8 s}
\left(1+g_H^2 \frac{g_V^2+g_A^2}{(1-m_Z^2/s)^2}-\frac{2 g_H g_V}{1-m_Z^2/s}\right) 
\sin^2\theta\label{31}
\end{eqnarray}
The total cross-section is:
\begin{equation}
\sigma_{tot}=\frac{\pi\alpha^2\kappa^3}{3 s}
\left(1+g_H^2 \frac{g_V^2+g_A^2}{(1-m_Z^2/s)^2} - 
\frac{2 g_H g_V}{1-m_Z^2/s}\right)\label{32}
\end{equation}
At this stage we note the characteristic angular 
distribution
of spin zero scalar particles which is expressed as:
\begin{eqnarray}
\frac{2 \pi}{\sigma_{tot}}\left(\frac{d \sigma}{d \Omega}\right)_0=
\frac{3 }{4}\sin^2\theta \nonumber 
\end{eqnarray}
This angular dependence leads to a vanishing forward--backward asymmetry $A_{FB}$ 
at tree-level,
where
$$ A_{FB} = \frac{\int_{\theta \leq \pi/2} d\Omega \frac{d\sigma}{d\Omega}  
- \int_{\theta \geq \pi/2} d\Omega \frac{d\sigma}{d\Omega} } 
{\int_{\theta \leq \pi/2} d\Omega \frac{d\sigma}{d\Omega}  
+\int_{\theta \geq \pi/2} d\Omega \frac{d\sigma}{d\Omega} } $$
 As we will see in the next section, at the one-loop level, 
only box diagrams can give contributions to  
$A_{FB}$.
\section{Radiative corrections.}
We have evaluated the radiative corrections at the one-loop level in
the 't Hooft--Feynman gauge. 
These one-loop corrections are ultraviolet (UV) 
and infrared (IR) divergent. 
The UV singularities are treated by  dimensional
regularization in the on--mass--shell renormalization scheme while the IR
singularities are regularized with a small fictitious photon mass $m_{\gamma}$. 
We renormalize not only the masses and coupling constants, but also the field
wave functions in such a way that the residues of renormalized
propagators are equal to one. 

The typical Feynman diagrams for the virtual corrections of order $\alpha^2$
are drawn in figure 1. These comprise, the THDM contribution to the photon 
and $Z$ propagator and their mixing (Fig.1.1), the standard 
contribution to the initial state (Fig. 1.2, 1.3 and 1.4), the THDM
contribution to the final vertex (Fig. 1.5--1.12) and the THDM box
contributions (Fig. 1.13 and 1.14).
The amplitudes of the Feynman diagrams depicted in Fig. 1.15 and Fig. 1.16 
are proportional to the electron mass and thus negligible. 
Bremsstrahlung diagrams and diagrams with a virtual photon emission
are shown in Fig 1.17--1.28. Diagrams 1.29--1.33 are needed for charged Higgs wave
function renormalization.

Note that  one loop contributions 
coming from initial state $e^+ e^- H_0$ and 
$e^+ e^- h_0$ vertices are also vanishing like $m_e$, since
$e^+$ and $e^-$ are both on-shell. A similar argument holds for the
$A^0$ and neutral goldstone exchange diagrams, as well as $Z-A^0$ and $Z-G^0$
mixing, noting also that $A^0 H^+ H^-$ and $G^0 H^+ H^-$ couplings
are already forbidden at tree-level by $CP$ invariance.

Those Feynman diagrams are generated and computed  using FeynArts and FeynCalc packages 
\cite{24} supplied with a
full MSSM and THDM-II Feynman rules code \cite{25}. We also used the 
Fortran FF-package
\cite{26}, in the numerical analysis. The one loop amplitude $ {\cal M}^1 $ 
projects fully on the two invariants defined in eq.(\ref{29}) as
\begin{equation}
 {\cal M}^1 =  {\cal M}_A^1 \ I_A +  {\cal M}_V^1 \ I_V \label{33}
\end{equation}
The typical one loop contributions are  given in this form in terms of the
Passarino--Veltman functions in appendix C and D.

The diagrams of figure 1: 1.22 and 1.25--1.28 together with real Bremsstrahlung 1.17--1.20 are needed to yield an IR finite corrected 
differential cross section (1.21, 1.23, 1.24 are free from IR divergencies) which has the following form:
\begin{equation}
\left(\frac{d \sigma}{d \Omega}\right)_1=\left(\frac{d \sigma}{d \Omega}\right)_0 +
 \left(\frac{d \sigma}{d \Omega}\right)_{SB} +2 Re({\cal M}_0^{*}{\cal M}^1 )
\frac{\kappa}{64 \pi^2 s}
\end{equation}
Here $\left(\frac{d \sigma}{d \Omega}\right)_{SB}$ denotes the soft 
Bremsstrahlung contribution.
Using eqs. (\ref{2.28}--\ref{30} and \ref{33}) the term $Re({\cal M}_0^{*}{\cal M}^1 )$
reads: 
\begin{equation}
Re({\cal M}_0^{*}{\cal M}^1 ) =  
[ {\cal M}_A^1 \Bigm ( -\frac{e^2 g_H g_A}{4(s-m_Z^2)}\ s^2 \ \kappa^2 \Big )
+ {\cal M}_V^1 \Bigm ( \frac{e^2}{4} \ \kappa^2 
(-s + s^2 \frac{g_H g_V}{(s-m_Z^2)} )\Big ) ]\sin^2\ \theta
\end{equation}
As one can see from the last formula the interference term is proportional to 
$\sin^2 \theta$. Consequently, since vertex   
and self--energy  contributions to  
${\cal M}_{A,V}^1$ have only $s$ dependence,
they will not contribute to the forward-backward asymmetry $A_{FB}$. 
Only box diagrams can contribute to $A_{FB}$ through their $t$ and $u$ 
dependence.

Let us cast the various one-loop corrections in the form: 
\begin{equation}
\sigma_1=\sigma_0 (1+\delta)=\sigma_0(1+\delta_{soft}^{\gamma} + 
\delta_{fermions}+
\delta_{bosons} )
\end{equation}
where $ \delta_{soft}^{\gamma}$ is the full (IR finite) set of one-loop QED contributions 
of diagrams 1.22--1.28 and of 1.17-1.21 with soft photon emission of figure 1, 
$\delta_{fermions}$ is the full vertex and self-energy contributions
of all leptons and quarks,
$\delta_{bosons}$ the massive gauge boson and Higgs contributions 
in vertices self-energies and boxes.  
\subsection{On--mass--shell Renormalization.}
The parameters entering the tree-level observables eqs.(\ref{31}--\ref{32}) are all
standard model parameters, except for the charged Higgs mass itself.
This fact will render the one-loop renormalization rather simple,
in the sense that all non-standard parameters appearing first at the
one-loop level, will not get renormalized. In particular, renormalization
conditions related to the definition of $\tan \beta$ are not explicitly 
needed here. We will need, however, to renormalize the charged Higgs
wave-function and mass. On the other hand, the renormalization scheme should 
be chosen in such a way to allow a simple interpolation between the MSSM and
the THDM-II. 
If we choose to renormalize $m_{A^0}$ on-shell, identifying the pole
of the propagator with the mass parameter in the renormalized Lagrangian
${\cal L_R}$, then $m_{H^\pm}$
would be uniquely determined in the MSSM, through the mass sum-rules,
and its renormalized quantity would be shifted from the corresponding mass
parameter in ${\cal L_R}$, \cite{20}. In the THDM-II case, one would then need to 
define, somewhat artificially, the renormalized $m_{H^\pm}$ also
to depart from the corresponding parameter in ${\cal L_R}$, though in such
a way that one recovers the MSSM situation when $\lambda_i \to \lambda_i^{MSSM}$. 
To avoid such complications, the simplest will be to take in the MSSM $m_{H^\pm}$ 
as a free parameter 
(rather than $m_{A^0}$) and renormalize it on-shell. This allows to treat
the MSSM and THDM-II cases at equal footing in the charged Higgs sector, 
with a unique renormalization scheme.
 
We will adopt throughout, the renormalization scheme of 
refs. \cite{Hollik}--\cite{Dabelstein-Hollik}. 
In this scheme one renormalizes all the 
fields before electroweak symmetry breaking. A 
renormalization constant $Z_{\Phi_{1,2}}$ is introduced for each doublet 
$\Phi_{1,2}$, $Z_2^{W}$ for the $SU_L(2)$ triplet $W_\mu^a$
and $Z_2^B$ for the $U(1)$ singlet. The gauge fields, coupling constants and 
vacuum expectation values $v_i$ are renormalized as follow:    
\begin{eqnarray}
& &W_\mu^a \rightarrow (Z_2^W)^{1/2} W_\mu^a \nonumber\\
& &B_\mu \rightarrow (Z_2^B)^{1/2} B_\mu \nonumber\\
& &\Phi_i \rightarrow (Z_{\Phi_i})^{1/2} \Phi_i \nonumber\\
& &g \rightarrow (Z_1^W)(Z_2^W)^{-3/2} g \nonumber\\
& &g' \rightarrow (Z_1^B)(Z_2^B)^{-3/2} g' \nonumber\\
& &v_i \rightarrow (Z_{\Phi_i})^{1/2} (v_i-\delta v_i)\label{38}
\end{eqnarray}
In the on--mass--shell scheme the counterterms can be fixed by the 
following renormalization conditions:
\begin{itemize}
\item The on-shell conditions for $m_W$, $m_Z$, $m_e$ 
and the electric charge $e$ are defined as in the standard 
model \cite{Hollik}.
\item On-shell condition for the charged Higgs boson $H^\pm$:
we choose to identify
the physical charged Higgs mass with the corresponding parameter in ${\cal L_R}$,
and require the residue of the propagator to have its tree-level 
value, i.e., 
\begin{equation}
\delta m^2_{H\pm} = Re \, {\sum }^{H^+H^+} (m^2_{H^\pm})\ \
and \ \ 
 \delta Z^{H\pm} = \frac{\partial}{\partial p^2}( 
{\sum}^{H^+H^+} (p^2)) |_{p^2=m^2_{H^\pm}} 
\end{equation}
where $\sum_{}^{H^+H^+} (p^2)$ is the charged Higgs bare self-energy.
\item Tadpoles are renormalized in such a way that the renormalized tadpoles
vanish: $T_{h} +\delta t_h=0$, $T_H +\delta t_H=0$. These conditions 
guarantee that $v_{1,2}$ appearing in ${\cal L_R}$ are located at the minimum
of the one-loop potential.
\end{itemize} 
As we are looking
for a charged Higgs bosons pair production and using $m_{H^\pm}$ and tan$\beta$
as independent parameters (in the Higgs sector),
we can choose a renormalization scheme very close to 
that of \cite{Dabelstein-Hollik}. The main difference with 
\cite{Dabelstein-Hollik} being that we renormalize the charged Higgs on-shell 
rather than the CP-odd $A_0$.

To compute the counterterms necessary to our study, we have to make
all the substitutions given in eqs.(\ref{38}) in the  covariant derivative and 
replace the renormalization constants $Z_i$ by their expansion up to first
order, $Z_i=1+\delta Z_i$. Those transformations shift 
$(D_{\mu}\Phi_i)^+(D_{\mu}\Phi_i)$ to the corresponding expression
in terms of renormalized fields and couplings plus a set of counterterms. 
In our case we need just the two following ones: 
\begin{eqnarray}
& &\delta (A_\mu H^+ H^-)=-i e [\delta Z^{H^{\pm}}+
(\delta Z_1^{\gamma}-\delta Z_2^{\gamma})
+g_H (\delta Z_1^{\gamma Z}-\delta Z_2^{\gamma Z}) ]
(k_1-k_2)_\mu\nonumber\\
& &\delta (Z_\mu H^+ H^-)=i e g_H [\delta Z^{H^{\pm}}+
(\delta Z_1^{Z}-\delta Z_2^{z})
+\frac{1}{g_H}(\delta Z_1^{\gamma Z}-\delta Z_2^{\gamma Z}) ]
(k_1-k_2)_\mu \label{40}
\end{eqnarray} 
where $\delta Z^{H^{\pm}} $ is related to $\delta Z_{\Phi_{1,2}}$ 
by the relation 
$\delta Z^{H^{\pm}}=\sin^2\beta \delta Z_{\Phi_1}+ \cos^2 \beta \delta Z_{\Phi_2}$,
the $\delta Z_i^{\gamma}$, $\delta Z_i^{z}$ and $\delta Z_i^{\gamma z}$ are 
related to $\delta Z_i^{W}$ and $\delta Z_i^{B}$ as in
\cite{Dabelstein-Hollik}\footnote{note ,however, a difference in the convention
for the sign of the Weinberg angle $\theta_W$.}. 
\begin{eqnarray}      
& & \delta Z_i^{\gamma}=s_W^2 \delta Z_i^W+c_W^2 \delta Z_i^B\nonumber\\
& & \delta Z_i^{Z}=c_W^2 \delta Z_i^W+s_W^2 \delta Z_i^B\nonumber\\
& & \delta Z_i^{\gamma Z}=-c_W s_W( \delta Z_i^W - \delta Z_i^B)=
\frac{-c_W s_W}{c_W^2-s_W^2}( \delta Z_i^Z - \delta Z_i^{\gamma})\nonumber
\end{eqnarray}
Moreover, in the on-shell scheme defined in ref
\cite{Hollik}, the wave function renormalization constants 
of the gauge bosons are given by: 
\begin{eqnarray}
& &\delta Z_2^{\gamma}=-\frac{\partial \Sigma^{\gamma}}{\partial p^2}(0)\\
& &\delta Z_1^{\gamma}=\delta
Z_2^{\gamma}+\frac{s_W}{c_W}\frac{\Sigma^{\gamma Z}(0)}{m_Z^2}\\
& &\delta Z_2^{Z}=\delta Z_2^{\gamma}+2\frac{c_W^2-s_W^2}{s_Wc_W}\frac{\Sigma^{\gamma
Z}(0)}{m_Z^2}+\frac{c_W^2-s_W^2}{s_W^2}(\frac{\delta m_Z^2}{m_Z^2}-
\frac{\delta m_W^2}{m_W^2})\\
& &\delta Z_1^{Z}=\delta Z_2^{\gamma}+\frac{3c_W^2-2s_W^2}{s_Wc_W}\frac{\Sigma^{\gamma
Z}(0)}{m_Z^2}+\frac{c_W^2-s_W^2}{s_W^2}(\frac{\delta m_Z^2}{m_Z^2}-
\frac{\delta m_W^2}{m_W^2})   \\
& &\delta Z_{1,2}^{\gamma, Z}=
-\frac{c_W s_W}{c_W^2-s_W^2} (\delta Z_{1,2}^Z-\delta Z_{1,2}^{\gamma})
\end{eqnarray}

Using these relations, one finds for the renormalization constants appearing 
in eq. (\ref{40}):  
\begin{eqnarray}
& & \delta Z_1^{\gamma} - \delta Z_2^{\gamma} =\frac{s_W}{c_W }
\frac{\Sigma^{\gamma Z}(0)}{m_Z^2} =  -\frac{\alpha}{2} 
B_0(0,m_W^2,m_W^2) \\ & &
\delta Z_1^{Z} - \delta Z_2^{Z} =\frac{c_W}{s_W}
\frac{\Sigma^{\gamma Z}(0)}{m_Z^2}= -\frac{\alpha c_W^2}{2 s_W^2 }
B_0(0,m_W^2,m_W^2) \\ &
& \delta Z_1^{\gamma Z} - \delta Z_2^{\gamma Z} =
-\frac{\Sigma^{\gamma Z}(0)}{m_Z^2}= \frac{\alpha c_W }{2 s_W } 
B_0(0,m_W^2,m_W^2) 
\end{eqnarray}
Note that the above combinations
are independent of the fermion and charged Higgs contributions because
$\Sigma^{\gamma Z}(p^2)$ is vanishing at $p^2=0$.\\
The renormalization constant of the charged Higgs wave function 
is given by:
\begin{equation}
\delta Z^{H^{\pm}}=\delta Z_{bosons}^{H^{\pm}} + \delta Z_{fermions}^{H^{\pm}}
\label{49}
\end{equation}
with:
\begin{eqnarray}
& & \delta Z_{bosons}^{H^{\pm}} = \frac{\alpha }{4 \pi s_W^2} ( 2s_W^2(
B_0(m_{\Hpm}^2,0,m_{\Hpm}^2)
+2 m_{\Hpm}^2 B_0'(m_{\Hpm}^2, m_\gamma^2,m_{\Hpm}^2) )
+ \nonumber\\ & & \frac{1}{2} B_0(m_{\Hpm}^2,m_A^2,m_{W}^2)
+\frac{c_{\beta\alpha}^2}{2} B_0(m_{\Hpm}^2,m_h^2,m_{W}^2) 
 +  \frac{s_{\beta\alpha}^2}{2} B_0(m_{\Hpm}^2,m_H^2,m_{W}^2)
+\nonumber\\ & & \frac{(-1+2s_W^2)^2}{2 c_W^2} B_0(m_{\Hpm}^2,m_{\Hpm}^2,m_{Z}^2)
- g_{hH^+H^-}^2
B_0'(m_{\Hpm}^2,m_h^2,m_{W}^2) +
\nonumber\\ & &
\frac{1}{4}(2 m_A^2+2 m_{\Hpm}^2-m_W^2 - 4 g_{AH^+G^-}^2 s_W^2)
B_0'(m_{\Hpm}^2,m_A^2,m_{W}^2) \nonumber\\ & &   
+\frac{1}{4}(c_{\beta\alpha}^2 (2 m_h^2+2  m_{\Hpm}^2- m_W^2) - 4
g_{hH^+G^-}^2 s_W^2) B_0'(m_{\Hpm}^2,m_h^2,m_{W}^2)\nonumber\\ & &
- g_{HH^+H^-}^2 B_0'(m_{\Hpm}^2,m_H^2,m_{W}^2)
 +\frac{1}{4}(s_{\beta\alpha}^2( 2 m_H^2+2
m_{\Hpm}^2-m_W^2)\nonumber\\ & &
- 4 g_{HH^+G^-}^2 s_W^2) B_0'(m_{\Hpm}^2,m_H^2,m_{W}^2) +
\frac{(4 m_{\Hpm}^2 -m_Z^2)}{4 c_W^2}(c_W^2 - s_W^2)^2
B_0'(m_{\Hpm}^2,m_{\Hpm}^2,m_{Z}^2) )\nonumber\\
& & \delta Z_{fermions}^{H^{\pm}} = \frac{N_C \alpha}{8\pi m_W^2 s_W^2} (
-(Y_u^2 + Y_d^2) B_0(m_{H^{\pm}}^2, m_d^2, m_u^2)  \nonumber\\ & &
 + ( (m_d^2 + m_u^2 - m_{H^{\pm}}^2 ) (Y_u^2 + Y_d^2) + 
4 m_d m_u Y_u Y_d ) B'_0(m_{H^{\pm}}^2, m_d^2, m_u^2) )\label{50}
\end{eqnarray}
where $\{c,s\}_{\beta\alpha}=\{\cos,\sin\}(\beta-\alpha)$,  
$Y_u$ and $Y_d$ are defined in eq. (\ref{27}).
 $N_C=3$ for quarks and 1 for leptons. 
The couplings $g_{AH^+G^-}$, $g_{hH^+H^-}$, $g_{HH^+H^-}$, $g_{hH^+G^-}$
and $g_{HH^+G^-}$ are trilinear scalar couplings which are model dependent.
$B_0'$ is the derivative of the $B_0$ function with respect to the square 
of the charged Higgs 4-momentum taken at $p_{H^\pm}^2= m_{H^\pm}^2$.  
In the above expressions, only $B_0'(m_{\Hpm}^2,m_{\gamma}^2,m_{\Hpm}^2)$ 
contains infrared divergences, regulated here by 
$m_{\gamma}$ which should cancel with the appropriate Bremsstrahlung 
contributions.

To sum up, most divergences (including initial state $\gamma e^+ e^-$ and 
$Z e^+ e^-$ vertices, and Z and $\gamma$ self-energies and mixing) are 
absorbed in the renormalization of the standard parameters,
the electric charge, $M_Z, M_W, m_e$ and the wave functions, as usual.
The one-loop correction to the $H^\pm$ self-energy
cancels out when the Higgs pair is produced on shell. 
The only remaining non-standard renormalization is that of $\gamma H^+ H^-$
and $Z H^+ H^-$ vertices, given by eqs. (\ref{40}). 
Furthermore, its model-dependence
is exclusively contained in the charged Higgs wave function renormalization
constant $\delta Z^{H^\pm}$ as can be seen from eqs. (\ref{49}, \ref{50}).
Hence the renormalization procedure will involve essentially
a set of (standard) parameters, which, combined with the quasi-susy
parameterization of section 2, would facilitate a simultaneous
treatment and comparison of THDM-II and MSSM cases.
\subsection{Bremsstrahlung}
In order to have an IR finite result we have to add to the cross section 
the contribution from  $e^+ e^-
\rightarrow H^+ H^- \gamma $ in the soft--photon limit; the
relevant diagrams are drawn in Fig. 1.17-1.28.
 
Denoting the photon four-momentum  by $k$ and its polarization vector
by $\epsilon_{\mu}(k)$ 
the amplitude of the soft Bremsstrahlung can be written in the form,
\begin{eqnarray}
\epsilon _{\mu}{\cal M}_{SB}^{\mu}&=& e {\cal M}_{Born} \left (
\frac{p_1\epsilon}{p_1k}-\frac{p_2\epsilon}{p_2k}+
\frac{k_2\epsilon}{k_2k}-\frac{k_1\epsilon}{k_1k} \right )
\nonumber\\
          & &-2 e^3 \bar v (p_2) \not \epsilon \left
(\frac{1}{s}+\frac{g_H(-g_V+g_A \gamma^5)}{s-m_z^2} \right )u(p_1) .\label{51}
\end{eqnarray}
The differential cross section is as follows:
\begin{eqnarray}
& &\Bigm ( \frac{d \sigma}{d \Omega}\Bigm ) _{_{{SB}}}=    \Bigm (
\frac{d \sigma}{d \Omega} \Bigm ) _{_0} \delta_{{\cal SB}}
\nonumber\\
& &+8 \frac{\alpha^3}{s}  ( 1-\frac{2 g_Vg_H}{1-m_Z^2/s}+
\frac{g_H^2 (g_V^2+g_A^2)}
{(1-m_Z^2/s)^2} )(I_0-\frac{s}{4}\kappa^2 \sin^2 \theta  (I_1+I_2))\label{52}
\end{eqnarray}

where $ \delta_{{SB}}$, $I_0$, $I_1$ and $I_2$ are given by, \\
\begin{eqnarray}
\delta _{{\cal SB}}&=&-\frac{\alpha}{\pi} \biggm \{ 2 ln\bigm (\frac{4 \Delta E^2}
 {m_{\gamma} ^2}\bigm )-ln\bigm (\frac{4 \Delta E^2}
{m_{\gamma} ^2}\bigm ) ln\bigm (\frac{s}{m_e ^2}\bigm ) +ln\bigm (\frac{m_e^2}{s}\bigm )
+\frac{1}{\kappa}ln(\frac{1-\kappa}{1+\kappa})  \nonumber\\ [0.3cm]
& &+\frac{1+\kappa ^2}{2 \kappa}
 ln\bigm (\frac{4 \Delta E^2}{m_{\gamma} ^2}\bigm )
 ln\bigm (\frac{1-\kappa }{1+\kappa }\bigm )
 +2 ln\bigm (\frac{4\Delta E^2}{m_{\gamma} ^2}\bigm ) ln\bigm (
 \frac{m_{H^+}^2-u}{m_{H^+}^2-t}\bigm )\nonumber\\ [0.3cm]
& &+\frac{\pi^2}{3}+\frac{1}{2} ln^2\bigm (\frac{m_e ^2}{s}\bigm )
+\frac{1+\kappa^2}{\kappa} \biggm [ Li_2\bigm (\frac{2
\kappa}{1+\kappa}\bigm )+\frac{1}{4} ln^2 (\frac{1-\kappa}{1+\kappa}) \biggm ] 
\nonumber\\ [0.3cm]
& & +2  \biggm [ Li_2\bigm (1-\frac{s(1-\kappa)}{2(m_{H^+}^2-t)}\bigm )+
 Li_2\bigm (1-\frac{s(1+\kappa)}{2(m_{H^+}^2-t)}\bigm )\nonumber\\ [0.3cm]
 & &-Li_2\bigm (1-\frac{s(1-\kappa)}{2(m_{H^+}^2-u)}\bigm )-
Li_2\bigm (1-\frac{s(1+\kappa)}{2(m_{H^+}^2-u)}\bigm ) \biggm ]  \biggm \}
\label{53} \\ [0.3cm]
I_0 &= &\kappa (\frac{\Delta E^2}{2}+\frac{\Delta E}{2\sqrt{s}}m_{H^\pm}^2)
+\frac{m_{H^\pm}^2}{2} (1-\frac{m_{H^\pm}^2}{s})
ln(\frac{(1-\kappa)^2}{(1-\kappa)^2+4\frac{\Delta E}{\sqrt{s}}\kappa})\nonumber 
\\ [0.3cm] 
I_1 & = & I_2=(1+\kappa^2) ln(\frac{(1-\kappa)^2}{(1-\kappa)^2+ 
4\frac{\Delta E}{\sqrt{s}}\kappa})
\end{eqnarray}
after integration over the photon momentum subject to $|\vec{k}| < \Delta E$,
$\Delta E$ defining the soft photon energy cut-off for the Bremsstrahlung process.
The infrared divergence is regulated by a fictitious small photon mass 
$m_{\gamma}$.
 
We have checked algebraicly and numerically that $m_{\gamma}$ cancels out
when the Bremsstrahlung cross-section is added to the rest. 
Note also that the Sudakov effects
$ln^2(m_e^2/s)$ cancel with the QED diagrams in the initial state. 
Numerically, we find that the second term of the right-hand side of eq. (\ref{52}) is very small compared to the term proportional to
$\delta_{\cal SB}$ and can be ignored. \\

Note finally that we assume throughout the study that hard bremsstrahlung
is suitably seperated. This relies of course on the quality of the veto on 
real photons in the final state.  

\section{Numerical analysis and discussion.}
 The following experimental input are taken for the physical parameters 
\cite{particledata}:
\begin{itemize}
\item the fine structure constant, $\alpha=\frac{e^2}{4\pi}=1/137.03598$.
\item the gauge boson masses, $m_Z=91.187\ GeV$, $m_W=80.41GeV$. 
\item the input lepton masses:
\begin{eqnarray}
& &m_e=0.511\ MeV \ \ \ \ \ \ \ \ \ \ \ \  m_{\mu}=0.1057\ GeV \ \ \ \ \ \ 
\ \ \ m_{\tau}=1.784 \ GeV  \nonumber
\end{eqnarray}
\item for the light quark masses we use the effective values which are 
chosen in 
such a way that the experimentally extracted hadronic part of the 
vacuum polarizations is reproduced 
\cite{martinzepenfieldVerzeganssijegerlener}:
\begin{eqnarray}
& &m_d=46\ MeV \ \ \ \ \ \ \ \ \ \ \ \ \ \ \ \ m_u=46 \ MeV \ \ \ \ \ \ \ \ \ \ \ \ 
\ \ m_s=150\ MeV \nonumber\\
& &m_c=1.5\ GeV \ \ \ \ \ \ \ \ \ \ \ \ \ \ \ \ m_b=4.5 \ GeV \nonumber
\end{eqnarray} 
\end{itemize}
The top quark mass is taken to be 
175 GeV. In the on-shell scheme we consider, $\sin^2 \theta_W$ is given by
$\sin^2 \theta_W\equiv 1- \frac{m_W^2}{m_Z^2}$ valid beyond tree-level.
The derived value $\sin^2 \theta_W \simeq 0.23$ thus includes radiative 
corrections.

Fig.2 shows the integrated tree level cross section $\sigma_0(fb)$ as a 
function of charged Higgs mass  for three values of $\sqrt{s}$. 
It can be seen that for a charged Higgs mass around $220$--$230\ GeV$ and for 
an integrated luminosity of about $10\ to \ 50 fb^{-1}$ (for the corresponding 
energy of 500 GeV), a few hundred events of charged Higgs pairs are expected 
at LC2000.  We can have the same situation for higher c.m. energies 
(1 TeV--2 TeV), for an integrated
luminosity  of about $200 fb^{-1}$, and for charged Higgs 
mass not very close to its threshold value.

We now discuss separately the various contributions to the quantum corrections. 
We have listed in table 1 the contribution to the  integrated cross
section (in percent) and to the forward--backward asymmetry due to the 
soft--photon
Bremsstrahlung and the virtual photon emission for several values of the 
soft--photon energy cut--off $\Delta E$. The dependence on  $\Delta E$ is 
rather important, for example, 
at $\sqrt{s}=500 GeV$ and for $m_{H^{\pm}} =220 GeV$ the correction increases
from $-10.8\% $ for $\Delta E= 0.15 E$ to reach $-24.7 \%$ at  
$\Delta E= 0.05 E$. The forward--backward asymmetry is small.  

Note that contributions from the initial $e^+e^-
\gamma, (Z)$ vertices to the corrected cross section 
are typically standard 
and hence depend neither on 
$\tan\beta$ nor on the charged Higgs mass. Those corrections
are of order (without soft and virtual photon emission)
 2.5\% at $\sqrt{s}=500$ GeV, 3.6\% at $\sqrt{s}=1000$ GeV and 
4.3\% at $\sqrt{s}=1500$ GeV.

The standard boxes
are not very sensitive to tan$\beta$, but can give important contributions.
For instance:
\begin{itemize} 
\item at $\sqrt{s}=500 GeV$ the
effect is about -9.5\% for a Higgs mass of $110\ GeV$ and  -7.3\% 
for a Higgs mass of
$220\ GeV$.
\item at $\sqrt{s}=1000 GeV$ the
effect is about -13.3\% for a Higgs mass of $230\ GeV$ and  -11.2\% 
for a Higgs mass of $410\ GeV$.
\item at $\sqrt{s}=1500 GeV$ the
effect is about -15.9\% for a Higgs mass of $410\ GeV$ and  -13.7\% 
for a Higgs mass of
$680\ GeV$.   
\end{itemize}

We discuss now the contribution of fermions (light fermions and top--bottom) 
to the
s--channel self--energy of the gauge bosons
 ($\gamma$--$\gamma$, $\gamma$--$Z$ and $Z$--$Z$) and
to the $\gamma H^+ H^-$ and $Z H^+ H^-$ vertices. 
Light fermion contributions are positive and  are not very sensitive  
to tan$\beta$ and $m_{H^{\pm}}$. As can be seen from table 2, these
contributions are substantial, around 16\%--19\%. 
The top-bottom  contribution is 
very sensitive to tan$\beta$ because of the form of the coupling given by 
eq.(\ref{27}) which depends on the top--mass. One can see from  eq.(\ref{27})
that the top effect is enhanced either for small values of $\tan\beta$
through the top mass effect, or for very large values of $\tan\beta$
through the non vanishing bottom mass. In such regimes effects of the
order of -25\% can be reached, thus cancelling or even overwhelming
the light fermion effects. Generically, however, the latter effects
remain dominant for intermediate $\tan \beta$ values.  

To discuss the contribution of the bosonic sector (Higgs bosons and gauge 
bosons
contribution) to the the gauge boson self energy, the
final state vertex and the box contributions
\footnote{Note that soft photon contributions 
have been subtracted and are given separately in
table 1.}, we will use the QSUSY parameterization described in section II.
 This parameterization  allows a comparison between the MSSM and
THDM-II and is described only by three parameters $m_{H^{\pm}}$, tan$\beta$ and
$\lambda_3$. \\
In Fig 3--5 we show the percentage contribution to the integrated cross section as a function of $\sqrt{s}$ and for different values of $\lambda_3$ and 
$\tan \beta$.
 For $m_{H^\pm}= 220$ GeV one finds that when $\lambda_3$ takes its 
supersymmetric value 
$\lambda_3^{MSSM}$ the
bosonic contribution is about -10\% at $\sqrt{s}\approx 500$ GeV 
(for both small and 
large tan$\beta$) and increases with increasing c.m. energy,
 reaching -20 \% to -25 \% for $\sqrt{s}\approx 2 TeV$. 
The correction is not very 
sensitive to
the charged Higgs mass and
interferes destructively with the fermionic contribution. 
In general, the bosonic contributions 
in the MSSM-like case tend to cancel the fermionic contributions
for not too small ($ < 1 $) or not too large ($\sim 60$) 
values of tan$\beta$.
When   
$\lambda_3$ is taken away from its supersymmetric value, the situation 
changes drastically. For 
a small deviation from $\lambda_3^{MSSM}$ the effect is still of 
the same order as in
the MSSM, but can get large and
have both  signs the farther we go from $\lambda_3^{MSSM}$, 
as can be seen from Fig.3,4,5. In some regions of the parameter space the 
effect can even grow too large for a relyable perturbative treatment,
reaching more than 40\% (!), without having enough cancellation
(if any) from the fermionic contribution. Such large effects can occur
as a direct consequence of increasing $\tan \beta$, as illustrated in
Fig. 6, but also
for small $\tan \beta$ as illustrated in Fig. 4.a, 5.a when
$m_{H^{\pm}}$ is large enough (see eqs.(22)). 

The total cross section $\sigma_1(fb)$ is plotted in figures 8 and 9 for 
two values of 
$m_{H^{\pm}}$ and tan$\beta$ and for various choices of $\lambda_3$. 
Shown is the summed effect of all fermionic 
and bosonic contributions (initial state, final state vertex, self energies
and non infrared boxes diagrams), excluding the gauge invariant set of diagrams 
containing at least one virtual photon exchange [the numerical contribution of 
the latter being mainly that of the IR sector].
When $\lambda_3=\lambda_3^{MSSM}$ the net effect of radiative corrections
remains small ($\lsim 5 \%$). In all cases we have 
illustrated, the one loop corrections generically decrease substantially the 
cross section for small $\tan \beta$ when $\lambda_3$ departs from  
$\lambda_3^{MSSM}$, while for large $\tan \beta$ the cross section can be
increased in some cases.  

Fig. 7 shows the one--loop forward--backward asymmetry $A_{FB}$ 
as a function of
$\sqrt{s}$. As one can see from this figure,
 $A_{FB}$ is small; this is due to the fact that  
only box diagrams contribute, and their couplings are $\lambda_3$ independent.\\

\section{Conclusion}
Future $e^+e^-$ linear colliders will probably offer the cleanest environment 
to discover a heavy charged Higgs through pair production. 
We have calculated the corresponding 
one loop radiative corrections in  
the on--mass--shell scheme, concentrating mainly here on a comparison between 
the 
contributions of the Higgs sector in a general THDM-II or MSSM-like contexts, 
as well as on the model-independent
soft photon contributions.
 We were lead to introduce a parameterization which 
allows a practical comparison between the two models and involves only one 
extra parameter (ex. $\lambda_3$) as compared to the MSSM.
The soft photon contribution for this process
is found to be substantial (about -20\%) for moderate values of
the soft--photon energy cut-off $\Delta E$.
If $\lambda_3$ takes its MSSM values
the one-loop corrections turn out to be rather small
(a few percent). 
Away from its MSSM values, $\lambda_3$ induces large effects, both for small and large 
tan$\beta$.  
Those large corrections arise  from the contributions of model--dependent 
vertices which enter both the final state vertex and the 
charged Higgs renormalization constant. 
We also illustrated the overall sensitivity to 
the charged Higgs mass, $\tan \beta$ and the c.m. energy.  

In summary, the radiative corrections to the charged Higgs pair production in 
$e^+ e^-$ annihilation can be significant and should be included in any 
reliable analysis. Furthermore, the large model-dependent effects can help
discriminate phenomenologically between a (softly broken) supersymmetric and a 
non-supersymmetric minimal Higgs sector, irrespective of direct evidence for 
the susy spectrum. The one-loop corrections of the full-fledged MSSM should
be eventually included. However they would not change much the conclusions
of this paper in as much as the supersymmetric partners remain
sufficiently heavy.\\

{\bf Acknowledgment:} We thank Michel Capdequi Peyran\`ere for his critical reading
of the manuscript and also for his collaboration in the early stage of this work. 
We are particularly indepted to Arndt Kraft for pointing out some errors
in previously published results and helping us correct them.
 
AA acknowledges a research grant from Cam\~oes Institut of Lisbonne and thanks 
the departemento de Fisica--Centra, Instituto Superior Tecnico for the warm 
hospitality during his visit where part of this work has been done.

\newpage
{\Large \bf Table Captions:}\\
\begin{itemize}
\item[{\bf Tab. 1:}] Soft photon contributions to the corrected cross section (in
percent) and to the forward--backward asymmetry with the following cut-off on 
the
photon energy:
$$\Delta E_1=0.05 \sqrt{s}/2\ \ , \ \ \Delta E_2=0.1 \sqrt{s}/2\ \ ,
\Delta E_3=0.15\sqrt{s}/2$$
\item[{\bf Tab. 2:}] Fermionic (leptons+light quarks+top--bottom)
 contributions to the corrected cross section (in percent) for $m_{top}=180\ GeV$.
\end{itemize} 
\vspace{0.4cm} 
{\Large \bf Figure Captions:}\\
\begin{itemize}
\item[{\bf Fig. 1:}] Feynman diagrams relevant for the 
${\cal O}(\alpha)$ contributions to 
$e^+ e^- \rightarrow H^+ H^- $. 1. 1) s--channel self energy diagrams, 1.2--1.4)
s--channel initial state vertex diagrams, 1.5--1.12) 
s--channel final state vertex diagrams, 1.13--1.14) THDM-II boxes diagrams, 
 1.17--1.21 Bremsstrahlung diagrams, 1.22--1.28)
diagrams with a virtual photon and 1.29--1.33) self energy of the charged Higgs bosons
necessary for the counterterms.
\item[{\bf Fig. 2:}] Integrated tree level cross section  as a function of $m_{H^{\pm}}$
for three value of $\sqrt{s}=500 GeV, \ 1000 GeV \ and \ 1500 GeV$.
\item[{\bf Fig. 3:}] Bosonic contribution (not including virtual and 
soft--photon emission)
to the integrated cross section as a function of $\sqrt{s}$ for  $m_{H^{\pm}}=220 GeV$,
3.a) tan$\beta=2$ and severals values for $\lambda_3$, 3.b) tan$\beta=30 $ and severals
values for $\lambda_3$
\item[{\bf Fig. 4:}] Same as in Fig.3 with $m_{H^{\pm}}=420 GeV$
\item[{\bf Fig. 5:}] Same as in Fig.3 with $m_{H^{\pm}}=730 GeV$
\item[{\bf Fig. 6:}] Bosonic contribution (in percent) to the integrated 
cross section as a
function of tan$\beta$, a)  $m_{H^{\pm}}=220 GeV$, $\sqrt{s}=500 GeV$ and different
 choices for $\lambda_3$ , b)  $m_{H^{\pm}}=300 GeV$, $\sqrt{s}= 1000 GeV$ and 
different choices 
for $\lambda_3$, c)  $m_{H^{\pm}}=420 GeV$ , $\sqrt{s}=1000GeV$...
\item[{\bf Fig. 7:}] Standard box contributions to forward--Backward asymmetry 
as function of $\sqrt{s}$ with
$m_{H^{\pm}}=220 GeV$, tan$\beta=2$ and for several values of $\lambda_3$ 
around $\lambda_3^{MSSM}$
\item[{\bf Fig. 8:}] Total one loop cross section $\sigma_1$ (in $fb$) as a
function of $\sqrt{s}$ for $m_{top}=180 GeV$ and $m_{H^{\pm}}=220 GeV$, 
8. a) tan$\beta=2$, 
$\lambda_3=\lambda_3^{MSSM}=-0.72$, $\lambda_3=-1.2$ and $\lambda_3=0.05$, 
8. b) tan$\beta=30$ , $\lambda_3=\lambda_3^{MSSM}=-0.72$ , $\lambda_3=-0.84$ 
and $\lambda_3=-0.8$.
\item[{\bf Fig. 9:}] Total one loop cross section $\sigma_1$ (in $fb$) as a
function of $\sqrt{s}$ for $m_{top}=180 GeV$ and $m_{H^{\pm}}=420 GeV$, 
9. a) tan$\beta=2$, 
$\lambda_3=\lambda_3^{MSSM}=-2.71$, $\lambda_3=-3.6$ and $\lambda_3=-1.8$, 
9. b) tan$\beta=30$ , $\lambda_3=\lambda_3^{MSSM}=-2.71$ , $\lambda_3=-2.9$ and 
$\lambda_3=-2.62$.
\end{itemize}
 
\newpage

\begin{center}
\begin{tabular}{|c|c|c|c|c|c|c|}  \hline
$\sqrt{s}$ & \multicolumn{2}{c|}{500GeV } & \multicolumn{2}{c|}{1 TeV} & 
\multicolumn{2}{c|}{1.5 TeV }\\ 
\hline
$ m_{H^+} GeV $   & 170 & 220   &  300 & 420 & 500 & 680      \\ \hline
  $\delta_{soft}^{\gamma}({\Delta E_1}) (\% )$   & -25.9 & -24.7   
&  -27.6 & -26.4 & -28.2 & -26.5     \\ \hline 
  $\delta_{soft}^{\gamma}(\Delta E_2) (\% )$   & -17. & -15.9   
&  -18.2 & -17.2 & -18.6 & -17.2     \\ \hline  
  $\delta_{soft}^{\gamma}(\Delta E_3) (\% )$   & -11.8 & -10.8   
&  -12.7 &  -11.9 & -12.9 & -11.7  \\ \hline 
  ${A_{FB}}_{soft}^{\gamma}(\Delta E_1) $   & 0.028 & 0.017   
&  0.033 & 0.021 &  0.031 & 0.0164   \\ \hline 
  ${A_{FB}}_{soft}^{\gamma} (\Delta E_2) $   & 0.0203 & 0.013   
&  0.024 &  0.0151 & 0.0223 & 0.012     \\ \hline  
  ${A_{FB}}_{soft}^{\gamma} (\Delta E_3)$   & 0.0166 & 0.01   
&  0.0193 &   0.0124 & 0.0183 & 0.0097    
   \\ \hline 
\end{tabular}
\end{center}
\centerline{{\bf Table 1.}}
\vspace{1.3cm}
\begin{center}
\begin{tabular}{|c|c|c|c|c|c|c|}  \hline
$\sqrt{s}$ & \multicolumn{2}{c|}{500GeV } & \multicolumn{2}{c|}{1 TeV} & 
\multicolumn{2}{c|}{1.5 TeV }\\ 
\hline
($ m_{H^+} (GeV)\ , tan\beta) $   & (170,0.7) & (220 ,0.7) &  (300, 0.7) & ( 420,0.7) & 
( 500,0.7) & (680,0.7)      \\ \hline
$\delta_{light-fermions}$ $(\%)$      & 17.     & 17    & 18.7   &  18.7   & 19.8
  &  19.8  \\ \hline 
$\delta_{top-bottom}$ $(\%)$    &-19.4  & -24.6  & -22.9   &  -16.5 & -17.5  & -13.6  
   \\ \hline  
$\delta_{total-fermions}$ $(\%)$       & -2.45   & -7.4   & -4.2  & 2.3   & 2.25   
& 6.2   \\ \hline
($ m_{H^+} (GeV)\ , tan\beta) $   & (170,2) & (220 ,2) &  (300, 2) & ( 420,2) & 
( 500,2) & (680,2)      \\ \hline
$\delta_{light-fermions}$ $(\%)$      & 17.     & 17    & 18.7   &  18.7   & 19.8
  &  19.8  \\ \hline 
$\delta_{top-bottom}$ $(\%)$    &-05.9  & -06.6  & -05.7   &  -04.9 & -04.7  & -04.2  
   \\ \hline  
$\delta_{total-fermions}$ $(\%)$       & 11.   & 10.3   & 13.  & 13.8   & 15.1   
& 15.6   \\ \hline 
($ m_{H^+} (GeV)\ , tan\beta) $   & (170,10) & (220 ,10)  & ( 300, 10) & (420,10) & 
(500,10) & (680,10)      \\ \hline  
$\delta_{light-fermions}$ $(\%)$  & 17.   & 17.    & 18.7   & 18.7   & 19.7 &
 19.7   \\ \hline  
$\delta_{top-bottom}$ $(\%)$    &-04.6  & -04.8   & -04.  & -03.7  & -03.4 & 
-03.3     \\ \hline 
$\delta_{total-fermions}$ $(\%)$ & 12.3    & 12.1  & 14.7   & 15.   & 16.3 
&  16.5    \\ \hline 
($ m_{H^+} (GeV)\ , tan\beta) $   & (170,20) & (220 ,20)  &  (300,20) & (420,20) 
& (500,20) & (680,20)      \\ \hline
$\delta_{light-fermions}$ $(\%)$      & 16.7     & 16.9   & 18.6 & 18.6 & 19.7  
& 19.7    \\ \hline 
$\delta_{top--bottom}$ $(\%)$      & -05.8  & -06.6  & -05.8 &   -05. & -04.8  
& -04.3\\ \hline  
$\delta_{total-fermions}$ $(\%)$      & 10.9 & 10.3   & 12.8  & 13.6  
& 14.9 &  15.4  \\ \hline 
($ m_{H^+} (GeV)\ , tan\beta) $   & (170,30) & (220 ,30)  &  (300, 30) & (420,30) 
& (500,30) & (680,30)      \\ \hline
$\delta_{ light-fermions}$ $(\%)$   & 16.8  & 16.8   & 18.5 & 18.6   &19.6     
& 19.6     \\ \hline  
$\delta_{top-bottom}$  $(\%)$   & -08. & -09.7    & -08.9 &-07.1   & -07.1 
& -06.  \\ \hline 
$\delta_{total-fermions}$ $(\%)$     & 08.8 & 07.1   &  09.6 & 11.5   
& 12.5 & 13.6     \\ \hline
($ m_{H^+} (GeV)\ , tan\beta) $   & (170,60) & (220 ,60)  &  (300, 60) & (420,60) 
& (500,60) & (680,60)      \\ \hline
$\delta_{ light-fermions}$ $(\%)$   & 16.2  & 16.3   & 17.8 & 18.1   &18.9     
& 19.2    \\ \hline  
$\delta_{top-bottom}$  $(\%)$   & -19.5 & -26.3    & -26. &-18.5   & -19.9 
& -15.3  \\ \hline 
$\delta_{total-fermions}$ $(\%)$     & -3.3 & -9.9   &  -8.1 & -0.04   
& -1. & 3.8     \\ \hline
\end{tabular}
\end{center}
\centerline{{\bf Table 2.}}


\newpage
\setlength{\baselineskip}{18pt}
\renewcommand{\theequation}{A.\arabic{equation}}
\setcounter{equation}{0}

\section*{Appendix}
\subsection*{Appendix A: Couplings}
For completeness, we give in this section the Feynman rules 
of the 3-point vertices involving the charged Higgs in the general THDM,
in two different forms.\\

In terms of the $\lambda_i$'s, $\alpha$ and $\beta$, one has

\begin{eqnarray}
g_{H^0 H^+H^-}&=&-i\sqrt{2} v[-\frac{\lambda_5}{2}\ \sin2\beta \ \sin(\alpha+\beta)\ +
\  (\lambda_4+2
\lambda_3)\ \cos(\beta-\alpha)\nonumber\\ & &\ \ \ \ \ \ \ \ \ \ \ +
 \sin2\beta \ (\lambda_2 \ \sin\alpha \ \cos\beta \ +\ \lambda_1 \ \cos\alpha 
\ \sin\beta)]\label{54}\\
g_{h^0 H^+H^-}&=&-i\sqrt{2} v[-\frac{\lambda_5}{2}\ \sin2\beta \ 
\cos(\alpha+\beta)\ 
+\  (\lambda_4+2
\lambda_3)\ \sin(\beta-\alpha)\nonumber\\ & &\ \ \ \ \ \ \ \ \ \ \ +
sin2\beta \ (\lambda_2 \ \cos\alpha \ \cos\beta \ -\ \lambda_1 \ \sin\alpha \ 
sin\beta)]\label{55}\\
g_{H^0 H^{\pm}G^{\mp}}&=&-\frac{iv}{\sqrt{2}} [ (\lambda_5-\lambda_4)\ \cos2\beta\
sin(\alpha+\beta)\ +\ \lambda_4\ \sin2\beta\ \cos(\alpha+\beta)\ \nonumber\\
& &\ \ \ \ \ \ \ \ \ \ \ +
2\ \sin 2\beta \ (\lambda_2 \ \sin\alpha \ \sin\beta \ -\ \lambda_1 \ \cos\alpha \ 
\cos\beta)] \nonumber\\
& &\ \ \ \ \ \ \ \ \ \ \ =\frac{-i g \sin(\beta-\alpha) (m_{H^\pm}^2-m_H^2)}{2 m_W}\\
g_{h^0 H^{\pm}G^{\mp}}&=&-\frac{iv}{\sqrt{2}} [ (\lambda_5-\lambda_4)\ \cos2\beta\
\cos(\alpha+\beta)\ -\ \lambda_4\ \sin 2\beta\ \sin(\alpha+\beta)\ \nonumber\\
& &\ \ \ \ \ \ \ \ \ \ \ +
2\ \sin2\beta \ (\lambda_2 \ \cos\alpha \ \sin\beta \ +\ \lambda_1 \ \sin\alpha \ 
\cos\beta)]\nonumber\\
& &\ \ \ \ \ \ \ \ \ \ \ =\frac{i g \cos(\beta-\alpha) (m_{H^\pm}^2-m_h^2)}{2 m_W}\\
g_{A^0 H^{\pm}G^{\mp}}&=& \mp \frac{v}{\sqrt{2}}\ (\lambda_6-\lambda_4)=
\mp \frac{m_{H^\pm}^2-m_A^2}{v \sqrt{2}}\label{58}
\end{eqnarray}
where the trigonometric functions of $\alpha$ should be expressed further in 
terms of the 
$\lambda_i$'s and $\tan \beta$, using eqs.(\ref{ABC},\ref{alph})
with $$v_1=\frac{v}{\sqrt{1+tan\beta^2}}\ \ \ \ \ \ \ \ \ \ \ \ \
v_2=v\sqrt{\frac{tan\beta^2}{(1+tan\beta^2)}}$$

Another useful form is in terms of deviations from the MSSM
tree-level mass-sum-rules. Defining

\begin{eqnarray}
&&\lambda_3=\frac{1}{8}(g^2+g'^2) - \lambda_1 + \frac{m^2_W}{v^2}\delta_3\\
&&m^2_{\Hpm}= (m^2_{\Hpm})_{MSSM} + m^2_W \delta_{\pm} \\
&&m^2_H = (m^2_H)_{MSSM} + m^2_W \delta_H \\
&&m^2_h = (m^2_h)_{MSSM} + m^2_W \delta_h
\end{eqnarray}
where $(m^2_{\Hpm})_{MSSM}$ and $(m^2_{H/h})_{MSSM}$ are given by 
eqs.(\ref{neutralmass}, \ref{chargedmass}), then


the Feynman rules for the vertices $H_0H^+H^-$ and
$h_0H^+H^-$ read
\begin{eqnarray}
&& g_{H^0H^+H^-}= g_{H^0H^+H^-}^{MSSM} -ig m_W [ \cos(\beta-\alpha)
(\delta_{\pm}
- \frac{\delta_H}{2}) +
\frac{ \sin(\alpha+\beta)}{ \sin2\beta
\tan^2\beta}\{ 4 \delta_3 -\frac{1}{2}(\delta_H+\delta_h)
 \nonumber \\ & & -\frac{1}{2 \cos^3\beta}(\cos (2\alpha + \beta) +
 \sin\beta (\frac{ \sin 2\alpha}{2} + \sin(\alpha-\beta) \cos(\alpha + \beta)))
(\delta_H - \delta_h)\}] \label{63} \\
&& g_{h^0H^+H^-}= g_{h^0H^+H^-}^{MSSM} -ig m_W [ \sin(\beta-\alpha)(\delta_{\pm}
- \frac{\delta_H}{2}) + 
 \frac{\cos(\alpha+\beta)}{ \sin2\beta
\tan^2\beta}\{ 4 \delta_3 -\frac{1}{2}(\delta_H+\delta_h)
\nonumber\\ & & - \frac{1}{2 \cos^3\beta}(\cos (2\alpha + \beta) + \sin\beta
(\frac{ \sin 2\alpha}{2} + \sin(\alpha+\beta) \cos(\alpha - \beta)))
(\delta_H - \delta_h)\}] \label{633}
\end{eqnarray}

It's interesting to note that in the general THDM the big
effects (in the case where tan$\beta$ is large or very small) comes
from the $\delta_3$ or $\delta_{H,h}$ but never from $\delta_{\pm}$.
Note also that these effects are not present not only in the susy case
but also when $\delta_H=\delta_h=4 \delta_3$.

\renewcommand{\theequation}{B.\arabic{equation}}
\setcounter{equation}{0}

\subsection*{Appendix B: Passarino--Veltman Functions}
Let us recall the definitions of scalar and tensor integrals we use:
The inverse of the propagators are denoted by
\begin{eqnarray} 
D_0= q^2-m_0^2 \ , \ D_i= (q+p_i)^2-m_i^2 \nonumber
\end{eqnarray}
Where the $p_i$ are the momentun of the external particles.\\
\\
{\bf One point functions:}
\begin{eqnarray} 
A_0(m_0^2)=\frac{1}{i \pi^2}\int d^nq \frac{1}{D_0} \nonumber
\end{eqnarray}
{\bf Two point functions:}
\begin{eqnarray} 
 B_{0,\mu}(p_1^2,m_0^2,m_1^2)=\frac{1}{i \pi^2}\int d^nq \frac{1,q_{\mu}}{D_0 D_1} 
\nonumber 
\end{eqnarray}
using Lorentz invariance, we have:
\begin{eqnarray} 
 B_\mu  =  {p_1}_{\mu} B_1 \nonumber 
\end{eqnarray}
{\bf Three point functions:}
\begin{eqnarray} 
 C_{0,\mu,\mu \nu} (p_1^2,p_{12}^2,p_2^2,m_0^2,m_1^2,m_2^2)=
\frac{1}{i \pi^2}\int d^nq \frac{ 1, q_{\mu},q_{\mu}q_{\nu}}{D_0 D_1 D_2} \nonumber
\end{eqnarray}
using Lorentz invariance, we have:
\begin{eqnarray}
& & C_\mu  =  {p_{1\mu} } C_1 +  p_{2\mu} C_2 \\ [0.4 cm]
& & C_{\mu\nu}  =  g_{\mu\nu} C_{00} + p_{1\mu} p_{1\nu} C_{11}
+ p_{2\mu} p_{2_\nu} C_{22} +
(p_{1\mu} p_{2_\nu} +p_{2\mu} p_{1_\nu} ) C_{12}
\end{eqnarray}
{\bf Four point functions:}
\begin{eqnarray}
 D_{0,\mu,\mu \nu} (p_1^2,p_{12}^2,p_{23}^2,p_3^2,p_2^2,p_{13}^2,m_0^2,
m_1^2,m_2^2,m_3^2)=
\frac{1}{i \pi^2}\int d^nq \frac{ 1, q_{\mu},q_{\mu}q_{\nu}}{D_0 D_1 D_2 D_3} 
\end{eqnarray}
using Lorentz invariance, we have:
\begin{eqnarray}
& & D_\mu  =  {p_1}_{\mu} D_1 +  p_{2\mu} D_2 +  p_{3\mu} D_3 \\ [0.4 cm]
& & D_{\mu\nu}  =  g_{\mu\nu} D_{00} + p_{1\mu} p_{1\nu} D_{11}
+ p_{2\mu} p_{2\nu} D_{22} + 
p_{3\mu} p_{3\nu} D_{33} +
(p_{1\mu} p_{2\nu} +p_{2\mu} p_{1\nu}) D_{12}\nonumber \\ [0.4 cm] & & +
( p_{1\mu} p_{3\nu} +p_{3\mu} p_{1\nu}) D_{13} +
(p_{3\mu} p_{2\nu} + p_{2\mu} p_{3\nu}) D_{23}
\end{eqnarray}

\renewcommand{\theequation}{C.\arabic{equation}}
\setcounter{equation}{0}
\subsection*{Appendix C: Vertex amplitudes. }
In this section we give only some typical amplitudes of the final state vertex
contributions drawn in figure 1.  The amplitude will be projected on the two
 invariants $I_V$ and $I_A$ defined in section 3. \\
The s--channel self-energy $\gamma$--$\gamma$, $\gamma$--$Z$ and $Z$--$Z$ 
can be found in 
\cite{Dabelsteinthesis}, \cite{katp}
The standard model initial state vertex contributions 
Fig 1.2, 1.3 and 1.4 are well known and will not be given here.\\ 

The remaining amplitudes are given as follow:

\subsection*{Diagram 1.5}
\begin{eqnarray}
 {\cal M}_{1.5} & = &\frac{\alpha^2
g_{SH^+W^-}g_{SH^-W^+}g_{VW^+W^-}}{ (s-m_V^2)}
\Bigm ( 
4\ B_0 (s,  m_W^2,  m_W^2) + (4 m_S^2 + s)\ C_0 -  
   4\ s\ C_{1}   \nonumber\\ & &  - 4\ C_{00} + 
       2\ s\ \kappa^2 ( C_{1 1} + C_{1 2} )
 \Bigm ) (g_V I_V-g_A I_A)
\end{eqnarray}
Where $g_{Z W^+W^-}=   \frac{c_W}{s_W}$, $g_{\gamma W^+ W^-}=1 $,
$g_{H_0 H^\mp W^\pm}=\pm  \frac{1}{2 s_W} s_{\beta\alpha}$, \\
$g_{h_0 H^\mp W^\pm}=\mp \frac{1}{2 s_W} c_{\beta\alpha}$,
$g_{A_0 H^\mp W^\pm}= \frac{- i}{2 s_W}$, with
$\{c,s\}_{\beta\alpha}=\{\cos,\sin\}(\beta-\alpha)$ and 
$\kappa ^2=1-\frac{\textstyle 4 m_{H^\pm}^2}{\textstyle s}$
and $m_S$ is the mass of the scalar particle S.\\
The $C_i$ and $C_{ij}$ have  as arguments $( m_{H^\pm}^2, s,  m_{H^\pm}^2, m_S^2,  m_W^2,  m_W^2)$.
\subsection*{Diagram 1.6}
$\bullet \ i\neq j$
\begin{eqnarray}
& & {\cal M}_{1.6} =  2\frac{\alpha^2
g_{VS_iS_j}g_{V'S_iH^-}g_{V'S_jH^+}}{(s-m_V^2)}
\Bigm ( (4\ m_{H^\pm} ^2 + m_{V'}^2 - 2\ s) C_0 + 4\ C_{0 0 } \nonumber \\ & &
       + (8\ m_{H^\pm} ^2 + m_{V'}^2 - 3\ s) ( C_{1} + C_{2} ) 
       - s\ \kappa^2 ( C_{1 1} + 2\  C_{1 2} +  C_{2 2} ) \Bigm )
(g_V I_V-g_A I_A)\label{69}
\end{eqnarray}
$\bullet \ i = j$
\begin{eqnarray}
& & {\cal M}_{1.6} =
2\frac{\alpha^2
g_{VS_iS_i}g_{V'S_iH^-}g_{V'S_iH^+}}{(s-m_V^2)}
\Bigm ( (4 m_{H^\pm}^2 + m_{V'}^2  - 2\ s) \ C_0 + 4 \ C_{00} + \nonumber \\ & &
             2 (8 \ m_{H^\pm}^2 + m_{V'}^2  - 3 \  s) C_{1} - 
            2\ s \kappa^2
              ( C_{11} +  C_{12} ) \Bigm ) (g_V I_V-g_A I_A)
\end{eqnarray}
The $C_i$ and $C_{ij}$ have as arguments $(m_{H^\pm}^2, s, m_{H^\pm}^2, {m_{V'}^2, 
m_{S_i}^2, m_{S_j}^2})$.
the coupling constants being defined in the following table. 
\begin{center}
\begin{tabular}{|c|c|c|c|c|}  \hline
 ($S_i$,$S_j$,$V'$ )    & ($H_{0}$, $A_0$, $W^+$) & ($h_{0}$, $A_0$,
$W^+$) &
 ($H^+$, $H^-$, $Z$) & ($ H^+$ , $H^-$, $\gamma$ ) \\ \hline
 $g_{Z S_i S_j} $   & $i\frac{s_{\beta\alpha}}{2s_W c_W}$   &
$  - i\frac{c_{\beta\alpha}}{2 s_W c_W}$
& $ - \frac{(c_W^2-s_W^2)}{2 s_W c_W}$    &
$ - \frac{(c_W^2-s_W^2)}{2s_W  c_W} $     \\ \hline
 $g_{\gamma S_i S_j} $   & 0  &
$ 0 $
& $ - 1$    &
$- 1 $     \\ \hline
 $g_{H^{\mp} S_i V'}  $   & $\pm \frac{1}{2 s_W} s_{\beta\alpha} $   &
$\mp \frac{1}{2 s_W} c_{\beta\alpha} $
& $-\frac{(c_W^2-s_W^2)}{2 s_W c_W}  $   & $ -  1  $   \\
\hline
 $g_{H^{\pm} S_j V'} $   & $ \frac{\mp i}{2 s_W}
$ & $ \frac{\mp i}{2 s_W} $
& $ -\frac{(c_W^2-s_W^2)}{2s_W c_W}    $& $- 1    $
   \\ \hline \end{tabular}
\end{center}

where $i, j$  label the five Higgs particles $h_0, H_0, A_0, H^+, H^-$
When V' is the photon, one has to keep consistently a photon mass
regulator in\\
 $C_0(m_{H^\pm}^2,m_{H^\pm}^2,s,m_{H^\pm}^2,m_{H^\pm}^2,m_{V'}^2)$
to account for the IR singularity, taking everywhere else in eq.(\ref{69})
$m_{V'} \to 0$. Summation over all $i,j$ should be understood for the full 
contribution of diagram 1.6. 

\subsection*{Diagram 1.7 and 1.8}
In this case:
if $V$ is the Z boson then $V'$ is the Z boson (resp. $W^\pm$ boson)
and $S_i$ is a neutral (resp. charged) Higgs boson  while $S_j$ is the
charged (resp. neutral ) one.\\
if $V$ is the photon then $V'$ is the $W^{\pm}$ gauge boson
and $S_i$ is a charged Higgs boson while $S_j$ is the neutral one.\\
 \begin{eqnarray}
& & {\cal M}_{1.7} = -\frac{\alpha^2
g_{S_iS_jH^+}g_{V'S_jH^-}g_{VV'S_i}}{(s-m_V^2)}
\Bigm ( C_0  
        - C_1  - C_2  \Bigm )
(g_V I_V-g_A I_A)
\end{eqnarray}
$C_i$ have as arguments 
$(m_{H^\pm}^2, s, m_{H^\pm}^2, m_{S_j}^2, m_{V'}^2, m_{S_i}^2)$

\begin{eqnarray}
& & {\cal M}_{1.8} = -\frac{\alpha^2
g_{S_iS_jH^+}g_{V'S_jH^-}g_{VV'S_i}}{(s-m_V^2)}
\Bigm ( C_0 - C_1  - C_2  \Bigm )
(g_V I_V-g_A I_A)
\end{eqnarray}
$C_i$ have as arguments $(m_{H^\pm}^2, s, m_{H^\pm}^2, m_{S_j}^2, 
m_{S_i}^2, m_{V'}^2)$\\
The coupling are given in the following table
\begin{center}
\begin{tabular}{|c|c|c|c|c|c|}  \hline
 (${S_i}$,$S_j$,$V'$ )    & ($H_{0}$, $H^+$, $Z$) & ($h_{0}$, $H^+$,
$Z$) &
 ($G^+$, $H_0$, $W^{\pm}$) & ($ G^+$, $h_0$, $W^{\pm}$)&($G^+$, $A_0$,
$W^{\pm}$)    \\ \hline
 $g_{Z V' {S_i}} $   & $ m_Z\frac{c_{\beta\alpha}}{s_W c_W}$   &
$m_Z\frac{s_{\beta\alpha}}{s_W c_W}$
& $-  m_Z s_W$    &
$-  m_Z s_W $ & $-  m_Z s_W $    \\ \hline
 $g_{\gamma V' {S_i}} $   & 0  &
$ 0 $
& $   m_W$    &
$  m_W $ & $  m_W $    \\ \hline
 $ g_{{S_i}S_jH^{\mp}}  $   & $-i g_{H_0 H^+H^-} $   & $-i g_{h_0H^+H^-} $
& $-i g_{G^+ H_0 H^- } $   & $-i g_{G^+ h_0 H^-}  $  & $ i g_{A^0 G^{\pm} 
H^{\mp} }$ \\
\hline
 $g_{V'S_jH^{\pm}} $   & $ -\frac{(c_W^2-s_W^2)}{2 s_W c_W}
$ & $ -\frac{(c_W^2-s_W^2)}{2 s_W c_W} $
& $\pm \frac{1}{2s_W} s_{\beta\alpha} $ & $\mp \frac{1}{2s_W} c_{\beta\alpha} $
 & $-i\frac{1}{2s_W}$  \\ \hline \end{tabular}
\end{center}
where $g_{H_0H^+H^-}$,$g_{h_0H^+H^-}$,$g_{H_0H^+G^-}$ and
 $g_{h_0H^+G^-}$ are defined in appendix A.
Summation over all $i,j$ should be understood for the full 
contribution of diagrams 1.7 and 1.8.
 
\subsection*{Diagram 1.9}
In the case of three different scalar $S_i$, $S_j$ and $S_k$ with
$S_i\neq S_j$, only the Z boson is coupled to two different scalar
(the coupling of the photon is forbidden because of the conservation of
the electromagnetic current )\\
$\bullet i\neq j$
\begin{eqnarray}
& & {\cal M}_{1.9} = 2\frac{\alpha^2
g_{H^+S_jS_k}g_{H^-S_iS_k}g_{VS_iS_j}}{(s-m_V^2)}
\Bigm ( C_0 + C_1 + C_2 \Bigm )
(g_V I_V-g_A I_A) \label{mijk}
\end{eqnarray}
$\bullet i = j$
\begin{eqnarray}
& & {\cal M}_{1.9} = 2\frac{\alpha^2
g_{H^+S_iS_k}g_{H^-S_iS_k}g_{VS_iS_i}}{(s-m_V^2)}
\Bigm ( C_0 + 2 C_1  \Bigm )
(g_V I_V-g_A I_A) \label{mijkp}
\end{eqnarray}
Where $C_i$ have as arguments: 
$( m_{H^\pm}^2 , s, m_{H^\pm}^2 , m_{S_k}^2 , m_{S_j}^2,
m_{S_i}^2 )$
\begin{center}
\begin{tabular}{|c|c|c|}  \hline
 ($S_i$,$S_j$,$S_k$ )    & ($H_{0}$, $A_0$, $G^+$) & ($h_{0}$, $A_0$,
$G^+$) \\ \hline
 $g_{Z S_i S_j} $   & $i\frac{s_{\beta\alpha}}{2 s_W c_W}$   &
$-i\frac{c_{\beta\alpha}}{2 s_W c_W} $    \\ \hline
 $g_{H^{\pm}S_j S_k} $   & $ i g_{A^0 G^{\mp} H^{\pm}} $   &
$ i g_{A^0 G^{\mp} H^{\pm}} $    \\ \hline
 $g_{H^-S_i S_k} $   & $-i g_{H_0H^-G^+}$   &
$-i g_{h_0H^-G^+} $    \\ \hline
\end{tabular} \end{center}

In this case, we have the following
situation: \\
If $S_i=H^+$ then $V=\gamma \ or \ Z$, $S_k=H_0 \ or \ h_0$
in this case we have $g_{Z H^+ H^-}=-\frac{ (c_W^2-s_W^2)}{2 s_W c_W}$,
$g_{\gamma H^+ H^-}=-1$.
The coupling $g_{H_0H^+H^-}$ and $g_{h_0H^+H^-}$
are model dependent and are given in eqs. (\ref{63},\ref{633}). \\

If $S_i=G^+$ then $V=\gamma \ or \ Z$, $S_k=H_0,h_0 \ or \ A_0$
in this case we have $g_{Z G^+ G^-}=-\frac{(c_W^2-s_W^2)}{2 s_W c_W}$,
$g_{\gamma G^+ G^-}=-1$. Summation over all $i,j$ should be understood for the 
full contribution of diagram 1.9.

\subsection*{Diagram 1.10 and 1.11}
The amplitude of diagram 1.10 is equal to the amplitude of 1.11.
 \begin{eqnarray}
& & {\cal M}_{1.10 + 1.11} = \frac{\alpha^2 g_{V V' S H^-}g_{S H^+ V'}
}{2m_{H^{\pm}}^2 (s-m_V^2)} ((m_S^2 - m_{V'}^2)B_0(0,m_S^2,m_{V'}^2 ) \nonumber\\ & &
+(- m_S^2 - 3 m_{H^{\pm}}^2 + m_{V'}^2) B_0 (m_{H^{\pm}}^2,m_S^2,m_{V'}^2) )
(g_V I_V-g_A I_A)
\end{eqnarray}
Where $V$ denotes the photon or the $Z$ gauge boson, 

$V'$ a charged (resp. neutral) gauge boson and $S$ a neutral (resp. charged)
scalar particle.
The couplings of these contributions are given by:
\begin{itemize}
\item  $W^-$ $H_0$ exchange:\\
$V=photon$,  $g_{\gamma W^- H_0 H^+}= -\frac{1}{2 s_W} s_{\beta\alpha} $,
$g_{H_0 H^\mp W^\pm}=\pm \frac{1}{2 s_W} s_{\beta\alpha}$\\
$V=Z$, $g_{Z W^+ H_0 H^-}= \frac{1}{2  c_W} s_{\beta\alpha}$ 
\item  $W^-$ $h_0$ exchange:\\ 
$V=photon$, $g_{\gamma W^+ h_0 H^-}= \frac{1}{2 s_W}c_{\beta\alpha} $,
$g_{h_0 H^\mp W^\pm}=\mp \frac{1}{2 s_W} c_{\beta\alpha}$\\ 
$V=Z$, $g_{Z W^+ h_0 H^-}= -\frac{1}{2  c_W} c_{\beta\alpha}$ 
\item  $W^-$ $A_0$ exchange:\\
$V=photon$, $g_{\gamma W^+ A_0 H^-}=\frac{-i}{2 s_W} $, 
$g_{A_0 H^\mp W^\pm}= \frac{-i}{2 s_W}$.\\
$V=Z$ , $g_{Z W^+ A_0 H^-}= \frac{i}{2  c_W} $ 
\item  $Z$ $H^+$ exchange:\\
$V=photon$, $g_{\gamma Z H^+H^-}=\frac{c_W^2-s_W^2}{s_W c_W}$  ,$g_{\gamma H^+ H^-}=-1$
\\
$V=Z$, $g_{Z Z H^+H^-}=\frac{(c_W^2-s_W^2)^2}{2 s_W^2 c_W^2}$,
$g_{ZH^+ H^-}=-\frac{c_W^2-s_W^2}{2s_W c_W}$ 
\item  photon $H^+$ exchange:\\
$V=photon$, $g_{\gamma\gamma H^+H^-}=2$, $g_{\gamma H^+ H^-}=-1$ \\
$V=Z$, $g_{\gamma Z H^+H^-}=\frac{c_W^2-s_W^2}{s_W c_W}$ , $g_{\gamma H^+ H^-}=-1$
\end{itemize}

\subsection*{Diagram 1.12}
For up-up-down exchange in the vertex the amplitude  is given by:
\begin{itemize}
\item  photon exchange:
\begin{eqnarray}
 & & {\cal M}_{1.12} =  
\frac{e_uN_C \alpha^2}{ m_W^2 s_W^2 s }
 \Bigm (2 ( ( m_d^2  + m_{H^{\pm}}^2  + m_u^2) ( Y_u^2 + Y_d^2 ) + 
4 m_d m_u Y_u Y_d )
        C_{1} + \nonumber \\ & &
     2  (m_d^2 (Y_u^2 +Y_d^2 ) + 2 m_u m_d Y_u Y_d ) 
        C_0 + (Y_u^2 + Y_d^2) B_0(s, m_u^2, m_u^2) + 
       \Bigm ) I_V 
\end{eqnarray}
\item  Z--exchange:
\begin{eqnarray}
& & {\cal M}_{1.12} =  \frac{ -  N_C \alpha^2}{2 c_W m_W^2 ( s - m_V^2 ) s_W^3} 
 \Bigm ( ( 2 e_u s_W^2 ( Y_u^2 + Y_d^2 ) - Y_d^2 ) B_0(s, m_u^2, m_u^2) + 
\nonumber \\ & &
  \{ 4 e_u s_W^2 ( (Y_u^2 + Y_d^2) m_d^2 + 2 m_u m_d Y_u Y_d )
   - 2 m_u m_d Y_u Y_d  - 2 m_d^2 Y_d^2 \}    
         C_0 \nonumber \\ & &
 - 2 \{ (m_u Y_u +m_d Y_d)^2  + m_{H^{\pm}}^2 Y_d^2 
   - 2 e_u s_W^2 (  ( Y_u^2 + Y_d^2) ( m_d^2  + 
            m_{H^{\pm}}^2  + m_u^2 ) \nonumber \\ & &  
+ 4  m_d m_u  Y_u Y_d ) \} 
  C_{1} \Bigm ) 
( g_V I_V - g_A I_A )
\end{eqnarray}
\end{itemize}
The $C_{i}$  have as arguments 
$(m_{H^{\pm}}^2, s, m_{H^{\pm}}^2, m_d^2, m_u^2, m_u^2) $
The contributions of the down--down--up vertex are obtained from the ones
above through the replacement:

$$ e_u \rightarrow -e_d \ , \ Y_u \rightarrow Y_d \ , \ Y_d \rightarrow Y_u \ , \
 m_u \rightarrow m_d \ , \ m_d \rightarrow m_u $$

\renewcommand{\theequation}{D.\arabic{equation}}
\setcounter{equation}{0}
\subsection*{Appendix D: Box amplitudes.} 
In this section we give the box amplitudes both for standard and 
supersymmetric cases.

\subsection*{THDM-II boxes}

We give the general amplitude for THDM-II boxes which are drawn in 
figure 1 ( 1.13, 1.14, 1.28, 1.29 and 1.30 )   with
exchange of two gauge boson one scalar and one fermion. 
We denote by $V_1$ and $V_2$ the two gauge bosons with masses $m_2$ and $m_4$, 
$S$ a scalar particle with mass
$m_1$ and $l$ a lepton  (electron or neutrino) with  mass $m_3$ of which we keep
trace only in the arguments of the Passarino-Veltman functions, and neglect
otherwise.
 
We will use the following notations for the coupling:
\begin{itemize}
\item gauge boson--two scalars: $V_{\mu} S_1 S_2= i e g_H (p-p')_{\mu}$
\item gauge boson--two fermions: $V_{\mu}^i f f'=
i e \gamma_{\mu}(g_{V_i}-g_{A_i} \gamma_5)$
\end{itemize}
we find then:
$${\cal M}_{box}=\alpha^2 g_{H_1}g_{H_2} ( G_V\ I_V - G_A\ I_A )\ INV$$
with :
 
$G_A= {  g_{A_2} }\,{  g_{V_1} } + {  g_{A_1} }\,{  g_{V_2} } $,
$G_V= {  g_{A_1} }\,{  g_{A_2} } + {  g_{V_1} }\,{  g_{V_2} } $\\
and 
\begin{eqnarray}
& &INV= 
 ((-4\ C_0(m_e^2, m_e^2, s, m_2^2, m_1^2, m_4^2) - 
       (3 m_{H\pm}^2 - t )\ D_0 + 
  2\ ( -t + m_{H\pm}^2 )\
        D_1 + \nonumber \\ & &
       ( -t + m_{H\pm}^2 ) D_2 + 
       2\ ( -t + m_{H\pm}^2 )\ D_3 + 
       2 \ D_{00} +  
       ( m_{H\pm}^2 + t )\ D_{11} + 2\ t \ D_{22}
+ \nonumber \\ & &  ( m_{H\pm}^2 +3 t)\ D_{12}
+ 2\ ( m_{H\pm}^2 + t )\ D_{13} 
+   (m_{H\pm}^2 +3 t)\ D_{23}
+  ( m_{H\pm}^2 + t )\ D_{33}))
\end{eqnarray}
The $D_0$, $D_{i}\ i=1,2,3$ and $D_{ij}$ have as arguments
$(m_{H^\pm}^2,m_e^2,m_e^2,m_{H^\pm}^2,t,s,m_1^2,m_2^2,m_3^2,m_4^2)$.

In the case of exchange of neutral gauge bosons in the box, we have to 
add the crossed boxes. The amplitude can be deduced from the above one by 
changing $t$ into $u$ and a global sign.

In the case of photon exchange in the box we have to regularize the I.R
 divergence by a small 
mass. we have checked numerically and analytically that the I.R 
divergence cancels out when adding the soft photon Bremsstrahlung. 
 
\newpage
\renewcommand{\thepage}{}

\centerline{\epsffile[0 0 334 664]{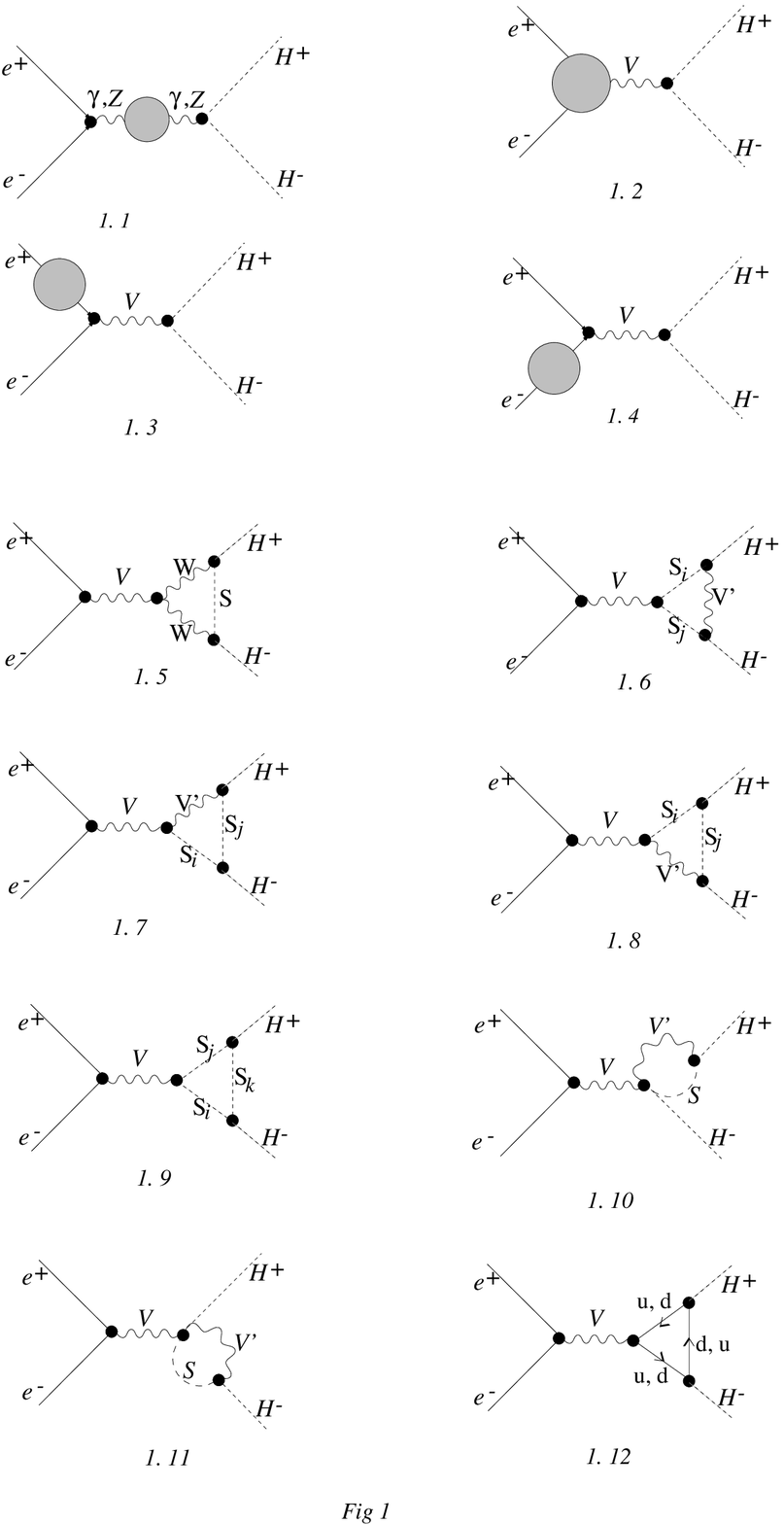} }
\newpage
\begin{minipage}[t]{19.cm}
\setlength{\unitlength}{1.in}
\begin{picture}(1.2,1)(0.8,8.2)
\centerline{\epsffile{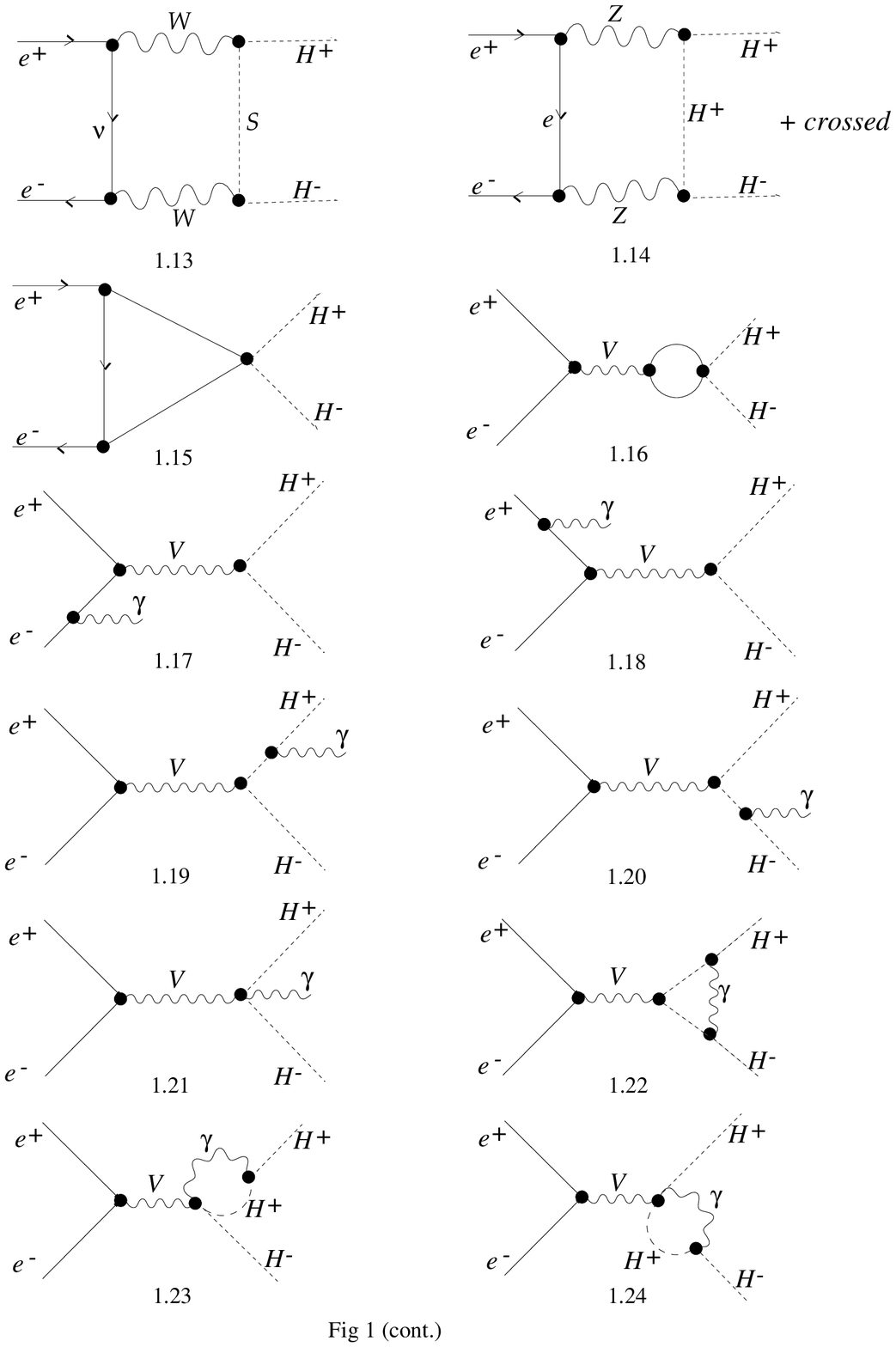}}
\end{picture}
\end{minipage}

\newpage
\begin{minipage}[t]{19.cm}
\setlength{\unitlength}{1.in}
\begin{picture}(0.1,0.1)(0.3,7.8)
\centerline{\epsffile{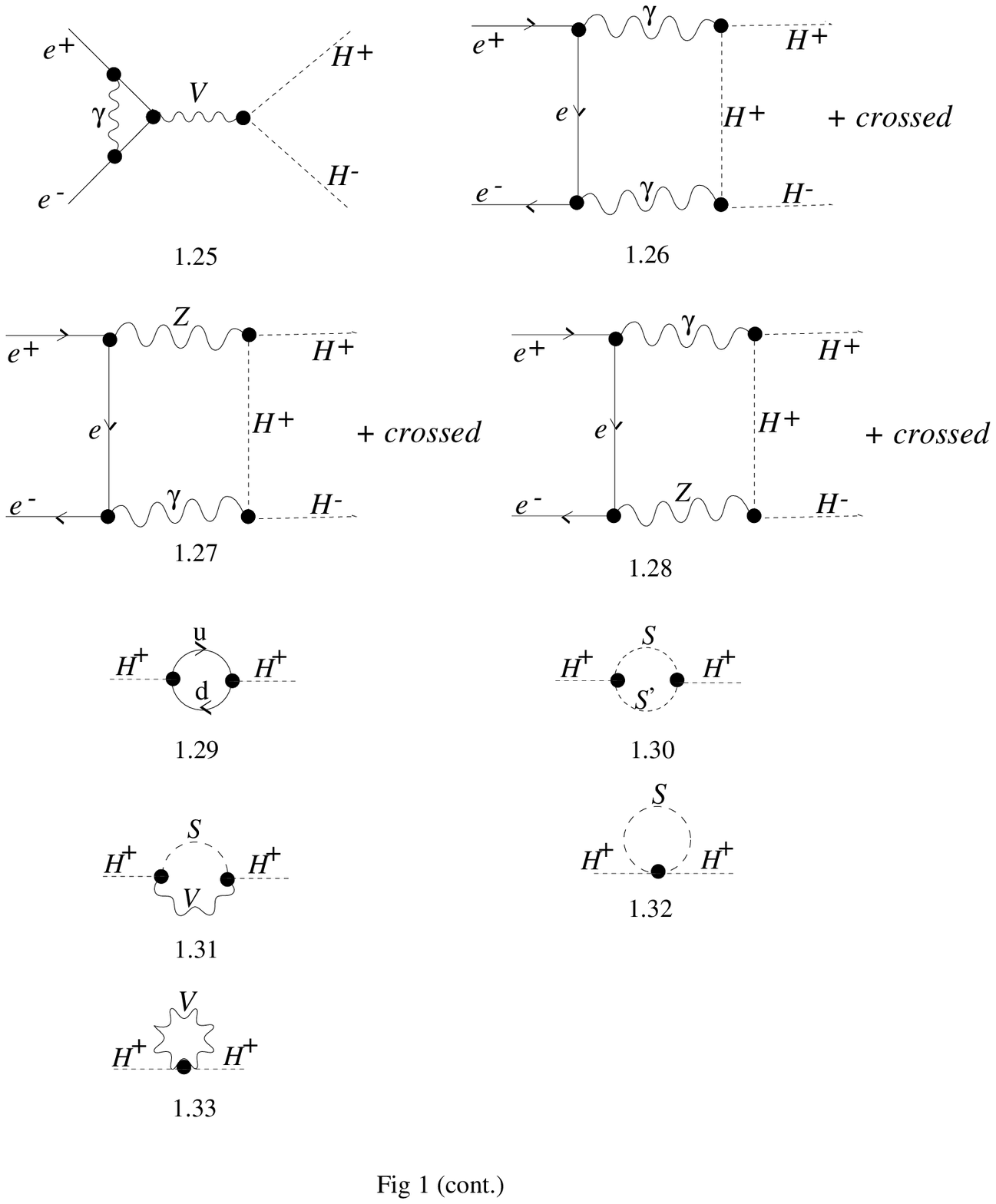}}
\end{picture}
\end{minipage}

\newpage
\begin{minipage}[t]{19.cm}
\setlength{\unitlength}{1.in}
\begin{picture}(1.2,1)(0.8,10.3)
\centerline{\epsffile{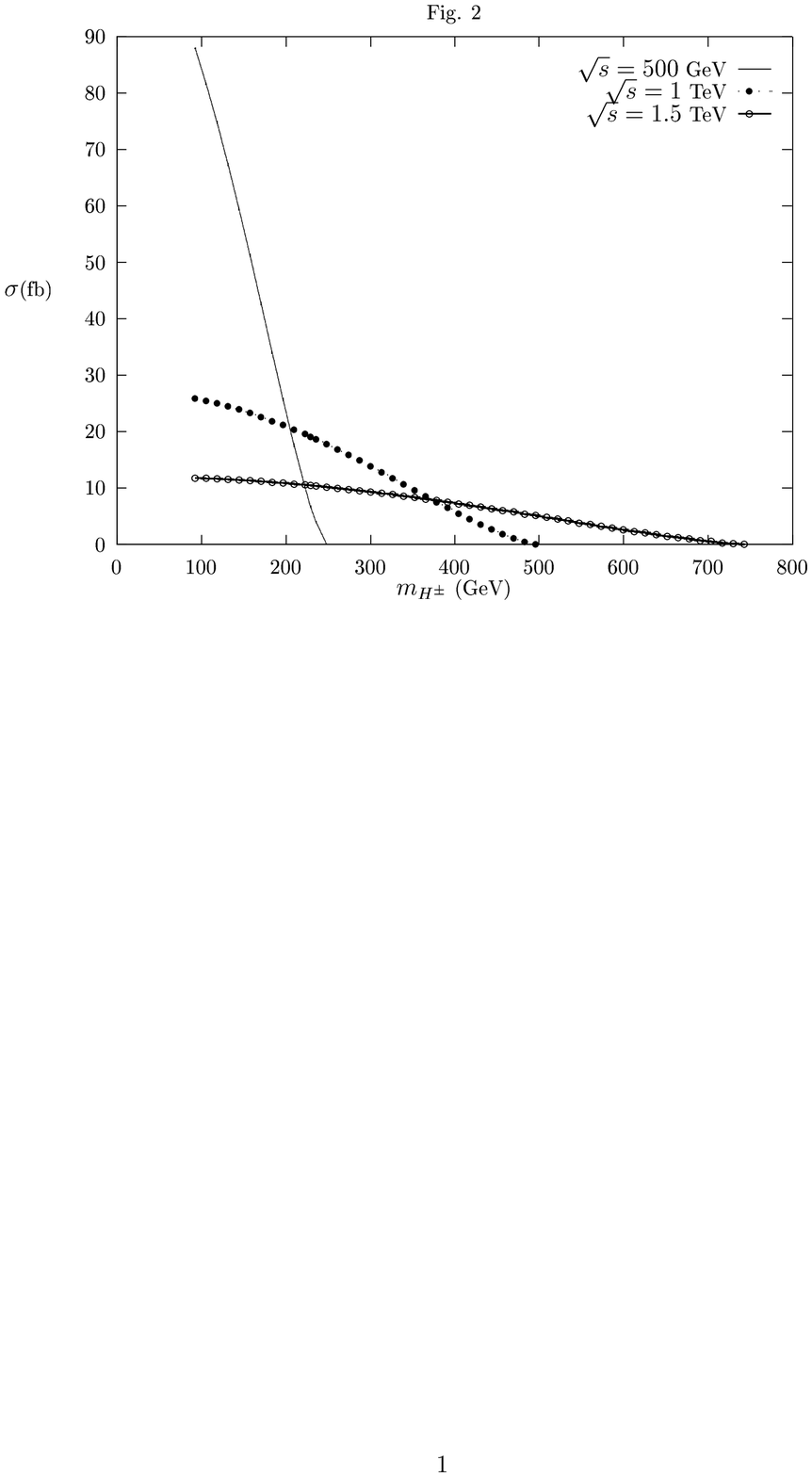}}
\end{picture}
\end{minipage}

\newpage
\begin{minipage}[t]{19.cm}
\setlength{\unitlength}{1.in}
\begin{picture}(1.2,1)(0.8,9.8)
\centerline{\epsffile{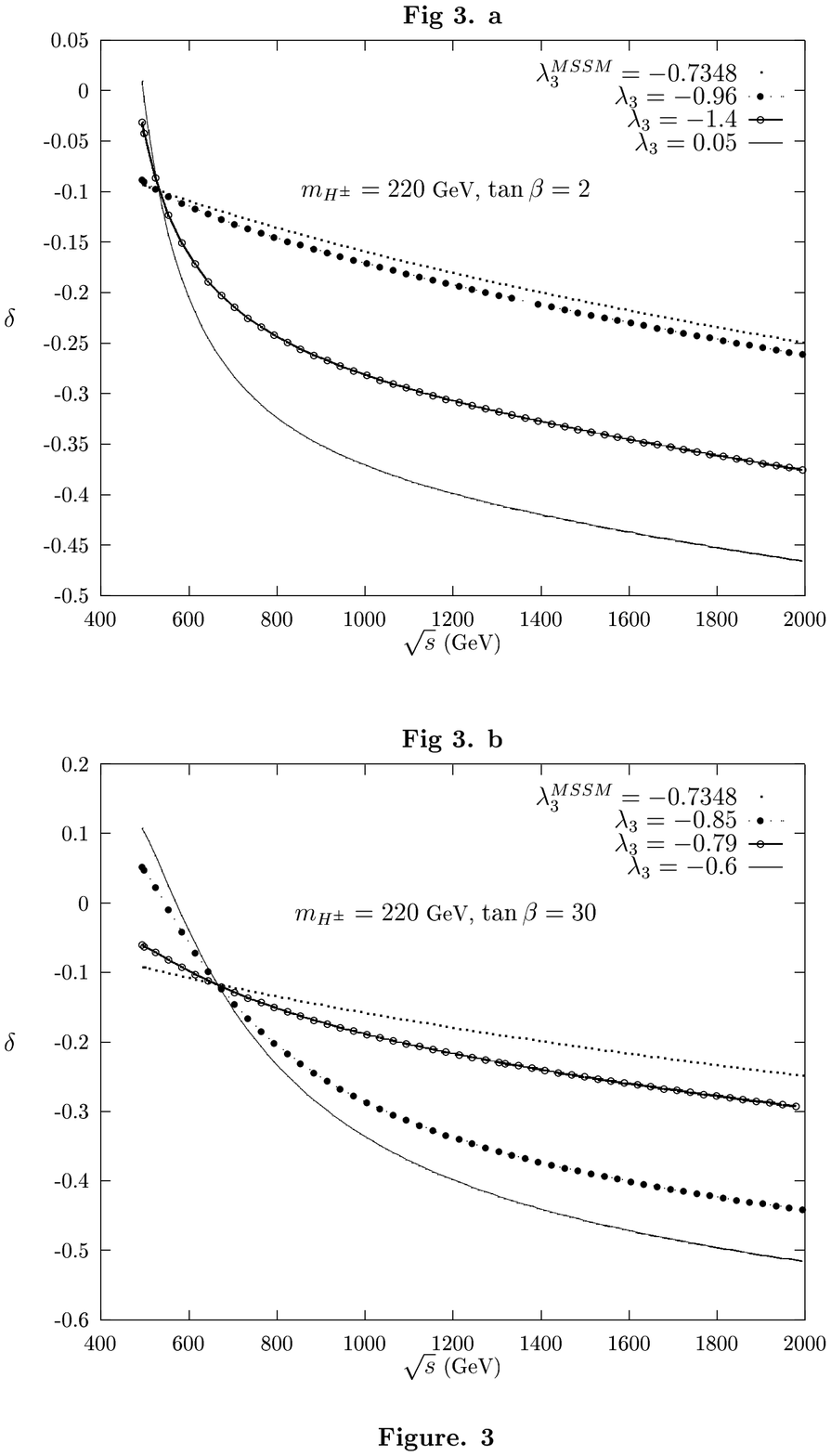}}
\end{picture}
\end{minipage}

\newpage
\begin{minipage}[t]{19.cm}
\setlength{\unitlength}{1.in}
\begin{picture}(1.2,1)(0.8,9.8)
\centerline{\epsffile{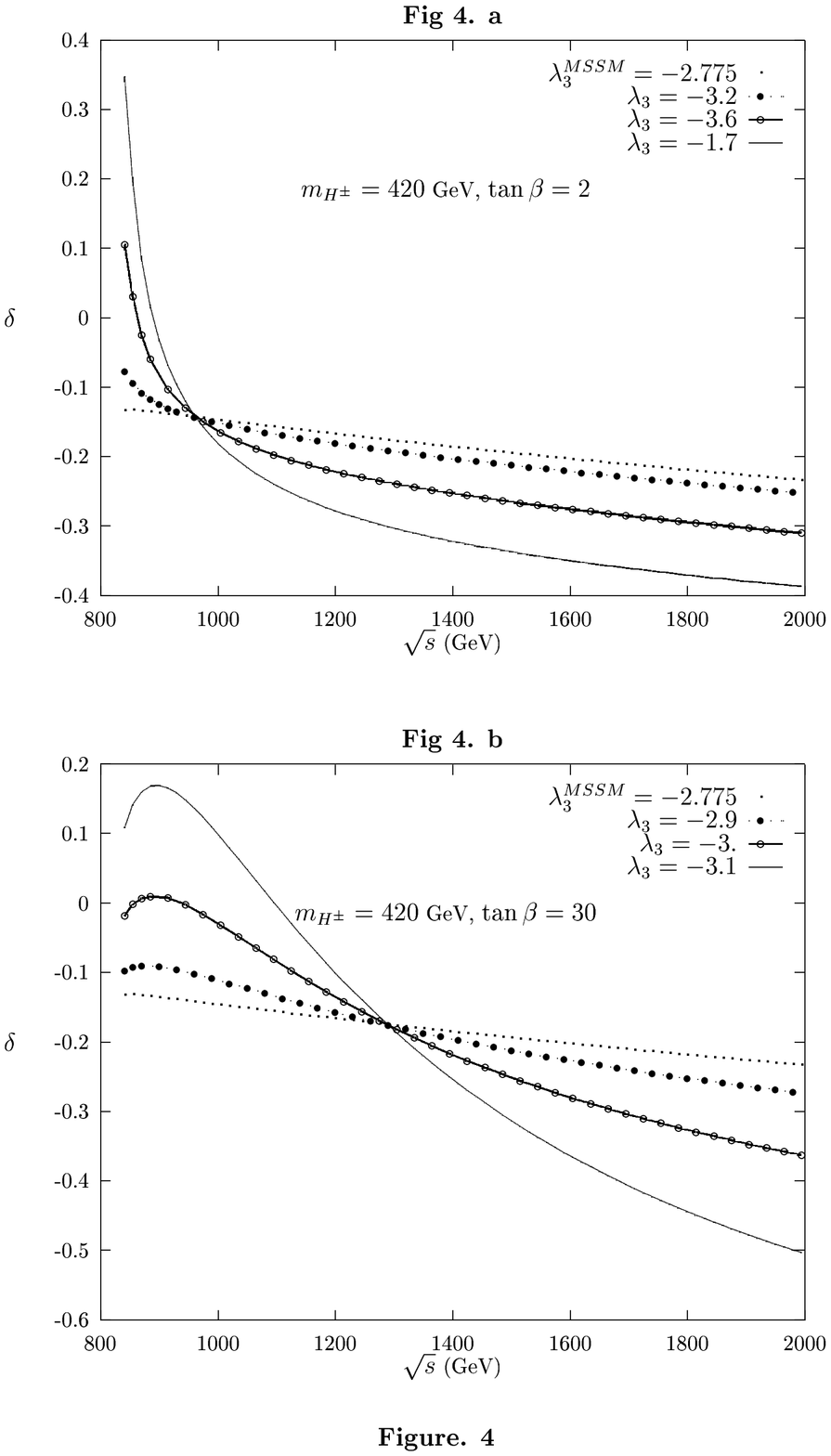}}
\end{picture}
\end{minipage}

\newpage
\begin{minipage}[t]{19.cm}
\setlength{\unitlength}{1.in}
\begin{picture}(1.2,1)(0.8,9.8)
\centerline{\epsffile{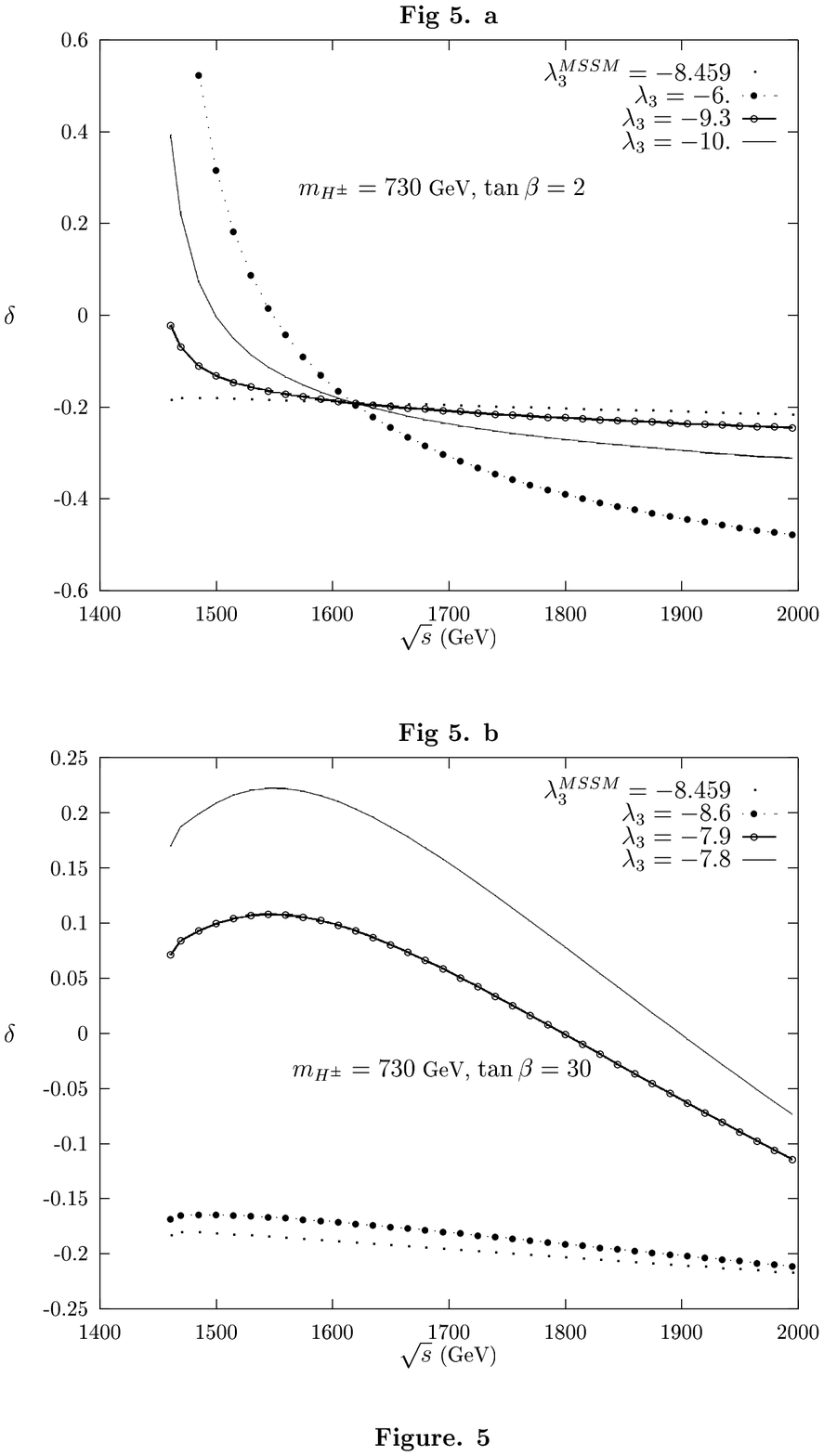}}
\end{picture}
\end{minipage}

\newpage
\begin{minipage}[t]{19.cm}
\setlength{\unitlength}{1.in}
\begin{picture}(1.2,1)(0.8,9.8)
\centerline{\epsffile{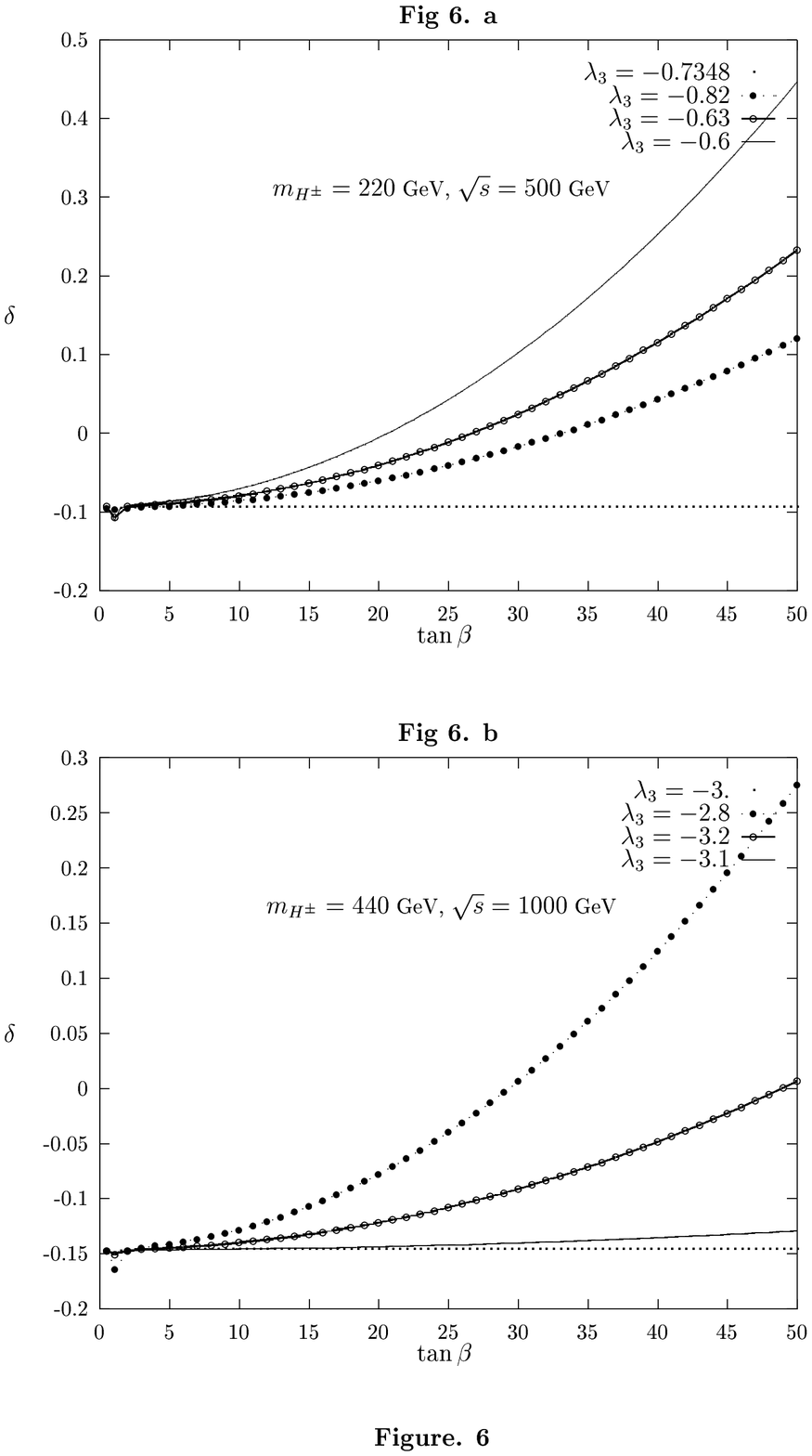}}
\end{picture}
\end{minipage}

\newpage
\begin{minipage}[t]{19.cm}
\setlength{\unitlength}{1.in}
\begin{picture}(1.2,1)(0.8,10.3)
\centerline{\epsffile{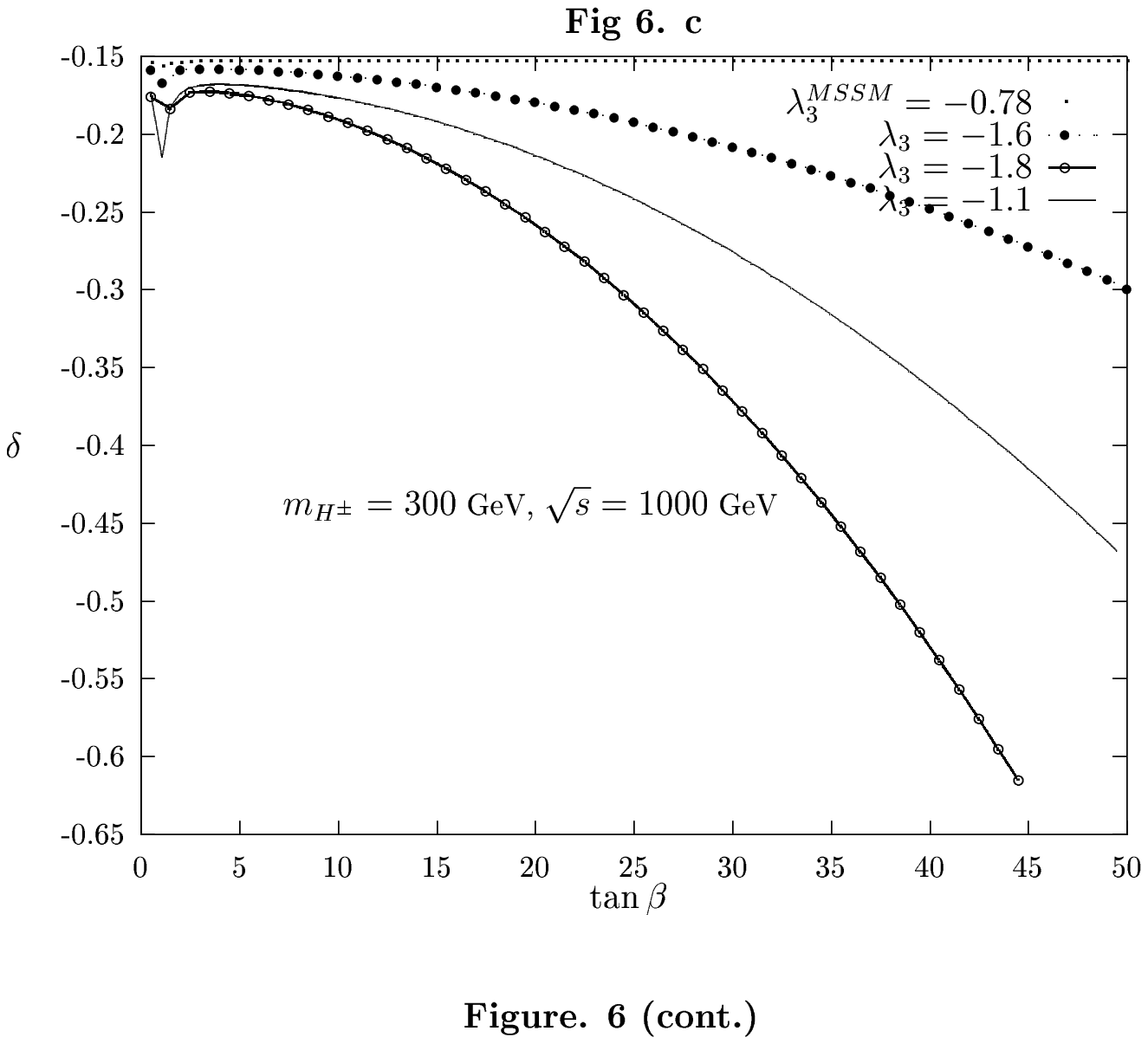}}
\end{picture}
\end{minipage}

\newpage
\begin{minipage}[t]{19.cm}
\setlength{\unitlength}{1.in}
\begin{picture}(1.2,1)(0.8,10.3)
\centerline{\epsffile{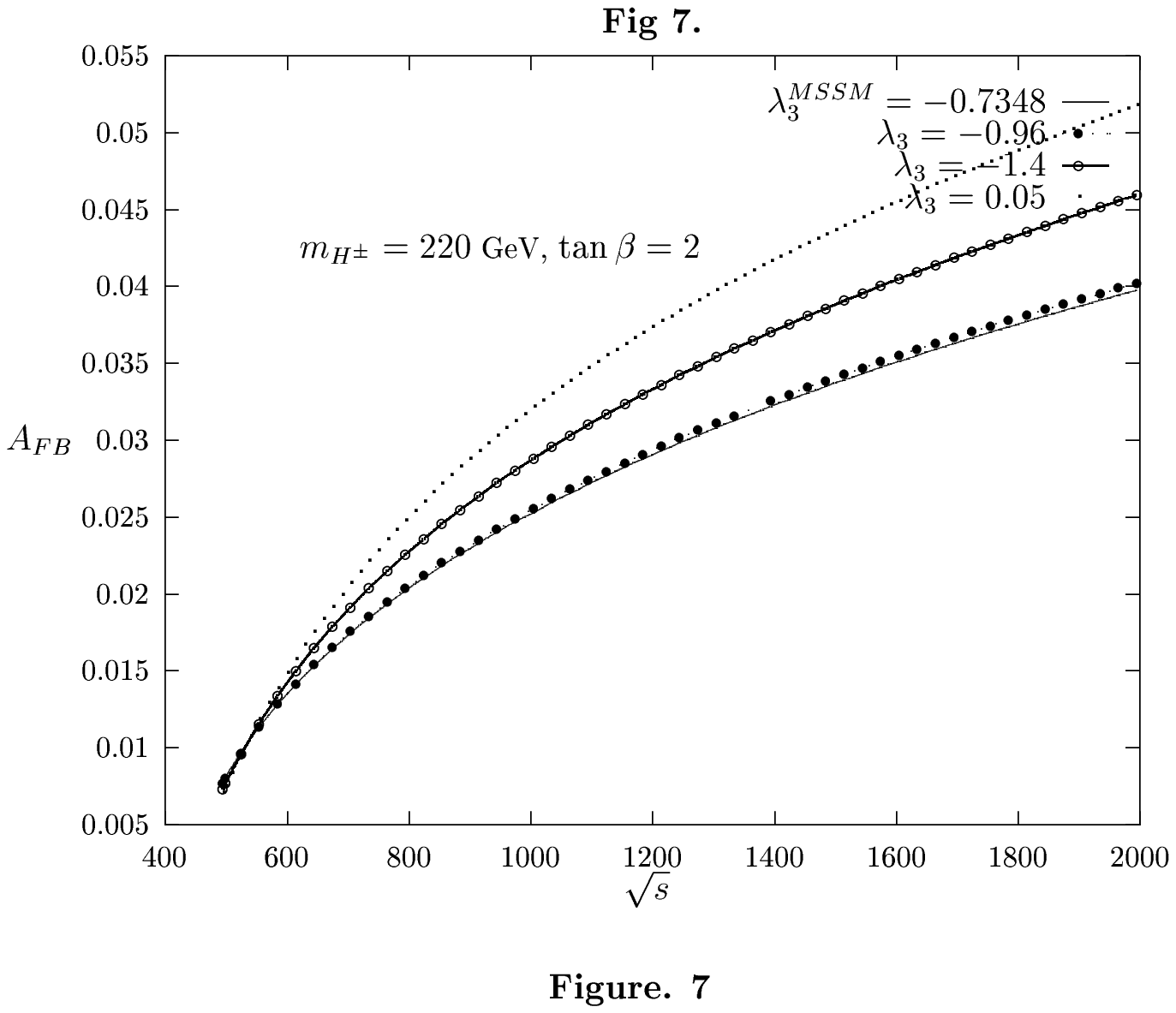}}
\end{picture}
\end{minipage}

\newpage
\begin{minipage}[t]{19.cm}
\setlength{\unitlength}{1.in}
\begin{picture}(1.2,1)(0.8,9.8)
\centerline{\epsffile{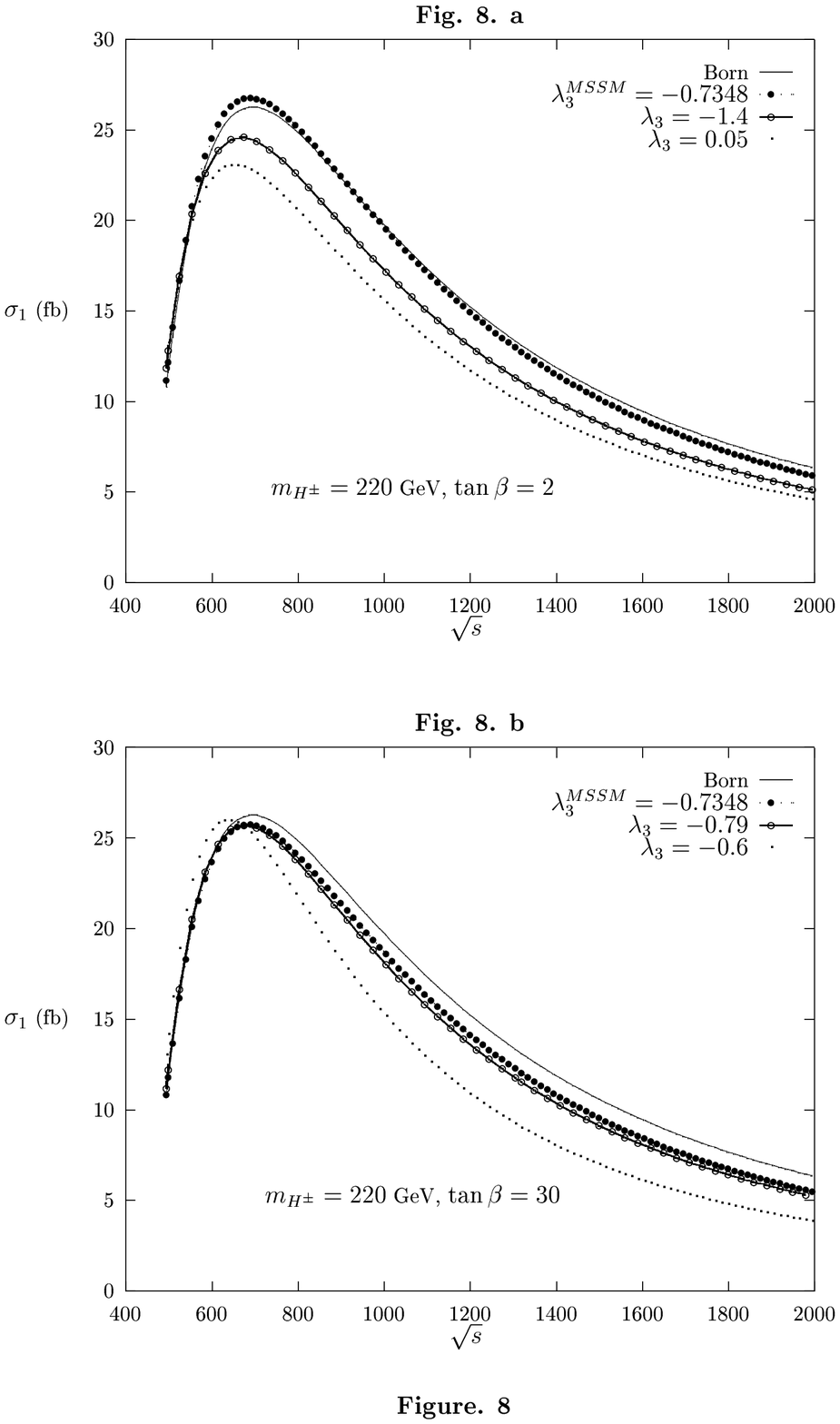}}
\end{picture}
\end{minipage}

\newpage
\begin{minipage}[t]{19.cm}
\setlength{\unitlength}{1.in}
\begin{picture}(1.2,1)(0.8,9.8)
\centerline{\epsffile{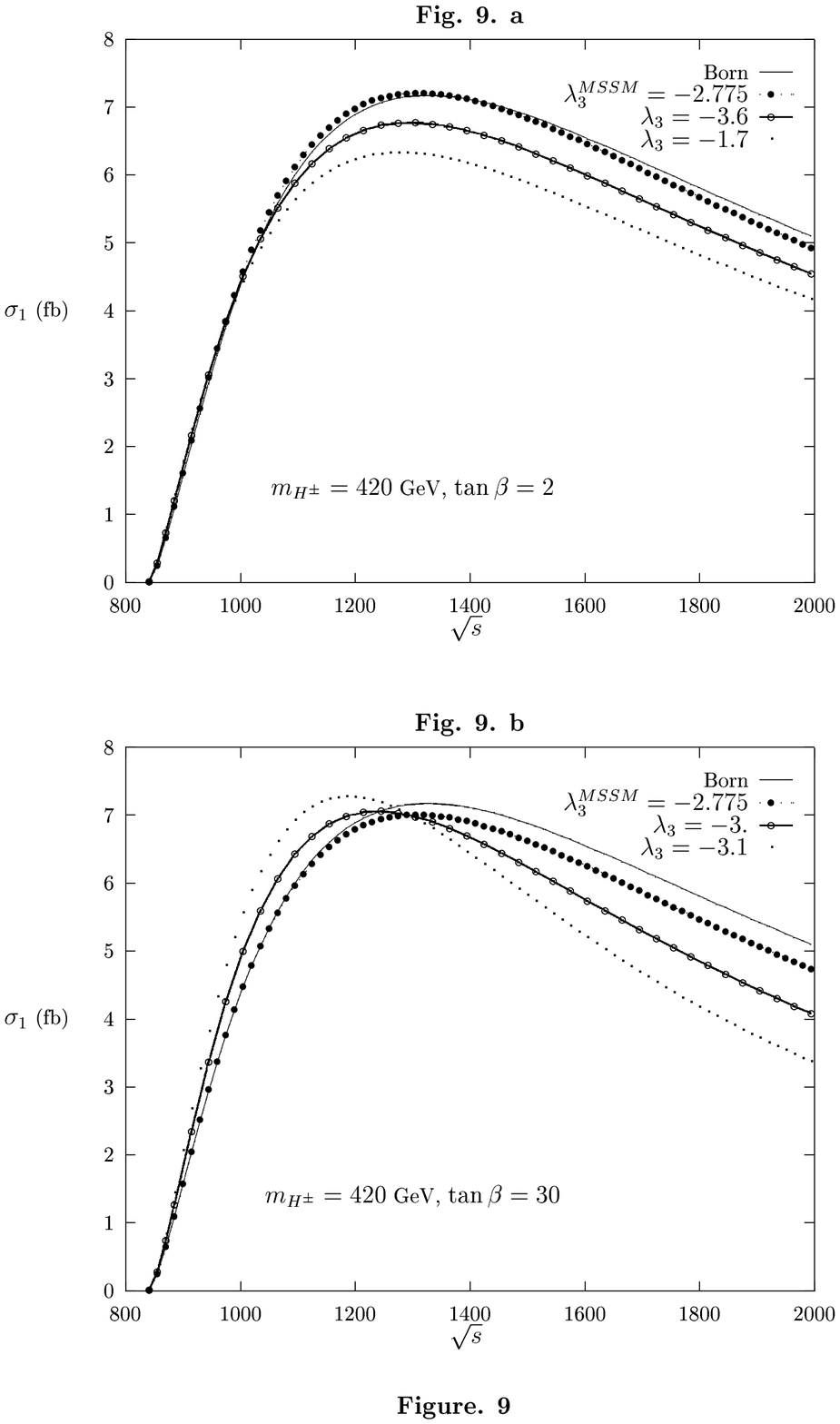}}
\end{picture}
\end{minipage}

\end{document}